\documentclass[journal]{IEEEtran}
\usepackage{subfigure}
\usepackage{epsfig} 
\usepackage{amsmath} 
\usepackage{amssymb}  
\usepackage{epstopdf}
\usepackage{subfigure}
\usepackage{cite}
\usepackage{balance}
\usepackage{color}
\usepackage{multirow}
\usepackage{booktabs}
\usepackage{algorithm}
\usepackage{algorithmicx}
\usepackage{algpseudocode}
\usepackage{multirow}
\usepackage{array}
\usepackage{hyperref}
\ifCLASSINFOpdf
\else
\fi

\begin{document}
\title{Electrical Load Forecasting Model Using Hybrid LSTM Neural Networks with Online Correction
}

\author{Nan~Lu,~Quan Ouyang,~\IEEEmembership{Member,~IEEE,}
Yang Li,~\IEEEmembership{Member,~IEEE,}~and  Changfu Zou,~\IEEEmembership{Senior Member,~IEEE}
\thanks{This work was supported by Marie Skłodowska-Curie Actions Postdoctoral Fellowships under the Horizon Europe programme (Grant No.~101067291). (Corresponding authors: Quan Ouyang and Changfu Zou)}

\thanks{Nan~Lu is with the College of Automation Engineering, Nanjing University of Aeronautics and Astronautics, Nanjing 211100, China, and also with the Faculty of Electrical and Electronic Engineering, University of Hong Kong, Hong Kong SAR 999077, China.(e-mail: nanlu@nuaa.edu.cn)}%
\thanks{Quan~Ouyang, Yang Li, and Changfu~Zou are with the Department of Electrical Engineering, Chalmers University of Technology, 41296 Gothenburg, Sweden. (e-mail:  quano@chalmers.se; yangli@ieee.org; changfu.zou@chalmers.se)}}

\markboth{}
{Shell \MakeLowercase{\textit{et al.}}: Bare Demo of IEEEtran.cls for IEEE Journals}

\maketitle

\begin{abstract}
 Accurate electrical load forecasting is of great importance for the efficient operation and control of modern power systems. In this work, a hybrid long short-term memory (LSTM)-based model with online correction is developed for day-ahead electrical load forecasting. Firstly, four types of features are extracted from the original electrical load dataset, including the historical time series, time index features, historical statistical features, and similarity features. Then, a hybrid LSTM-based electrical load forecasting model is designed, where an LSTM neural network block and a fully-connected neural network block are integrated that can model both temporal features (historical time series) and non-temporal features (the rest features). A gradient regularization-based offline training algorithm and an output layer parameter fine-tuning-based online model correction method are developed to enhance the model's capabilities to defend against disturbance and adapt to the latest load data distribution, thus improving the forecasting accuracy. At last, extensive experiments are carried out to validate the effectiveness of the proposed electrical load forecasting strategy with superior accuracy compared with commonly used forecasting models.

\end{abstract}


\IEEEpeerreviewmaketitle

\section{INTRODUCTION}

\IEEEPARstart{A}{ccurate} electrical load forecasting plays an indispensable role in the optimal planning and efficient operation of smart grids. It is a fundamental tool for improving the overall economy and stability of modern power systems while minimizing associated waste and costs. Even a $1\%$ reduction in the average forecast error can translate into hundreds of thousands of dollars in saving \cite{hobbs1999analysis}. With the rapidly growing market penetration of renewable energy and electric vehicles, and their integration into the grid, the power system is facing increasing volatility and complexity, thus bringing heightened challenges to accurate electrical load forecasting. 

A systematic review of the electrical load forecasting strategies was presented in \cite{KUSTER2017257}, where they are mainly classified into three categories: time series analysis, artificial neural network (ANN), and support vector regression (SVR). Commonly used time series analysis models include the auto-regressive integrated moving average (ARIMA) \cite{amjady2001short}, quantile regression averaging \cite{7137662}, etc. Although these models have the advantages of a simple structure and fast training speed, they cannot effectively reflect the volatility and nonlinearity in the electrical load time series, resulting in imprecise or unreliable forecasting outcomes. 

The ANN and SVR models utilize neural networks and classifiers, respectively, to capture the high-dimensional and complicated nonlinear correlation between the input features and output load, thus achieving higher forecasting accuracy \cite{wang2020short}. 
A deep neural network-based short-term load forecasting model was proposed in \cite{en11010213} based on the historical load data sequence, where three convolutional layers and three pooling layers were employed for feature extraction. In \cite{7748604}, an SVR-based short-term load forecasting algorithm was designed, where a two-step hybrid parameter optimization method was utilized to improve the forecasting accuracy.

As an improvement, some non-temporal features, such as the time index and similarity features, were added to the input for electrical load forecasting. For example, the Euclidean norm with weighted factors was adopted in \cite{982201} as a feature to evaluate the similarity between the forecast day and a searched previous day. Then, a one-hour-ahead load forecasting method was developed by using the correction of similar day data. Based on the past data sequence and the binary variable that specifies whether it is a working day, Jurado {\it et al.} proposed a hybrid methodology that combines feature selection based on entropies with soft computing and machine learning approaches \cite{JURADO2015276}. These authors showed that the forecasting errors can be decreased by adding this binary feature to the inputs. A long short-term memory (LSTM) neural network was developed in \cite{kong2017short} for short-term load forecast by taking into account the historical data sequence and its corresponding time of day indices, day of week indices, and binary holiday marks as the inputs. 
To further enhance prediction accuracy, a hybrid prophet-LSTM model optimized by back-propagation was designed in \cite{BASHIR20221678} for short-term electricity load forecasting.

Despite the significant progress represented by the above-mentioned approaches, none of them comprehensively incorporates all the features of historical data sequence, time index, and similarity. This may restrict the forecasting accuracy of the electrical load. Moreover, as the distribution of electrical load data changes with time, the forecasting error of the aforementioned models, obtained through offline training with historical information, tends to progressively increase. Considering these identified research gaps, we are therefore strongly incentivized to develop a hybrid LSTM neural network model with the capability of online correction for accurate day-ahead electrical load forecasting. 

The main contributions of this paper are summarized as:

\begin{enumerate}

 \item Four types of features are comprehensively incorporated for electrical load forecasting modeling, which include the historical time series, time index features, historical statistical features, and similarity features. Compared with the methods that only consider part of features, more accurate forecasting results can be achieved.
  \item A hybrid LSTM-based electrical load forecasting model is developed here, where an LSTM neural network block and a fully-connected neural network (FCNN) block are integrated that can model both temporal and non-temporal features, resulting in higher accuracy than the existing electrical load forecasting methods.
  
  \item A gradient regularization-based offline training strategy and an online correction strategy based on output layer parameter fine-tuning are adopted in this work, which can improve the anti-disturbance capability and calibrate the developed forecasting model to fit the latest load data distribution, thus further improving the forecasting performance.
\end{enumerate}

\section{PROBLEM STATEMENT}
\subsection{Dataset Analysis}
\begin{figure}[!htbp]
	\centering		
	\subfigure[]{
		\begin{minipage} {0.9\linewidth}\label{bel}
			\centering
			\includegraphics[width=1\linewidth, clip]{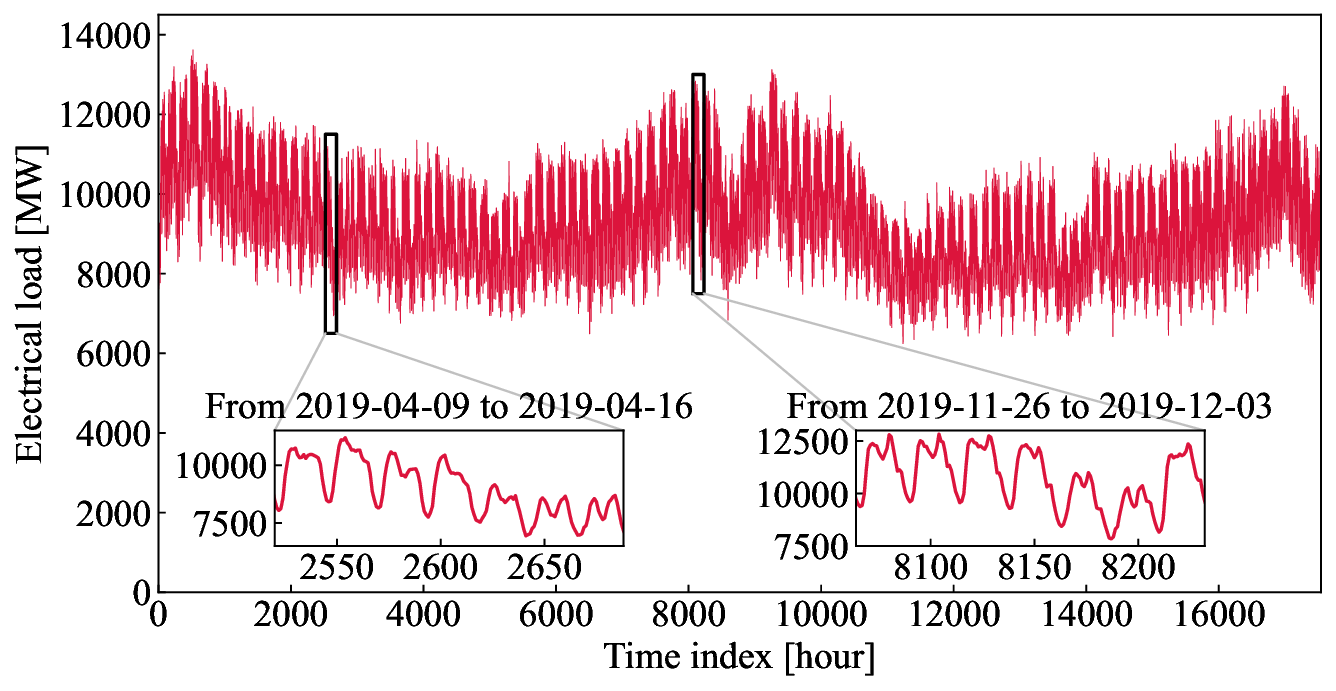}
		\end{minipage}
	}	
	\subfigure[]{
		\begin{minipage} {0.9\linewidth}\label{den}
			\centering
			\includegraphics[width=1\linewidth]{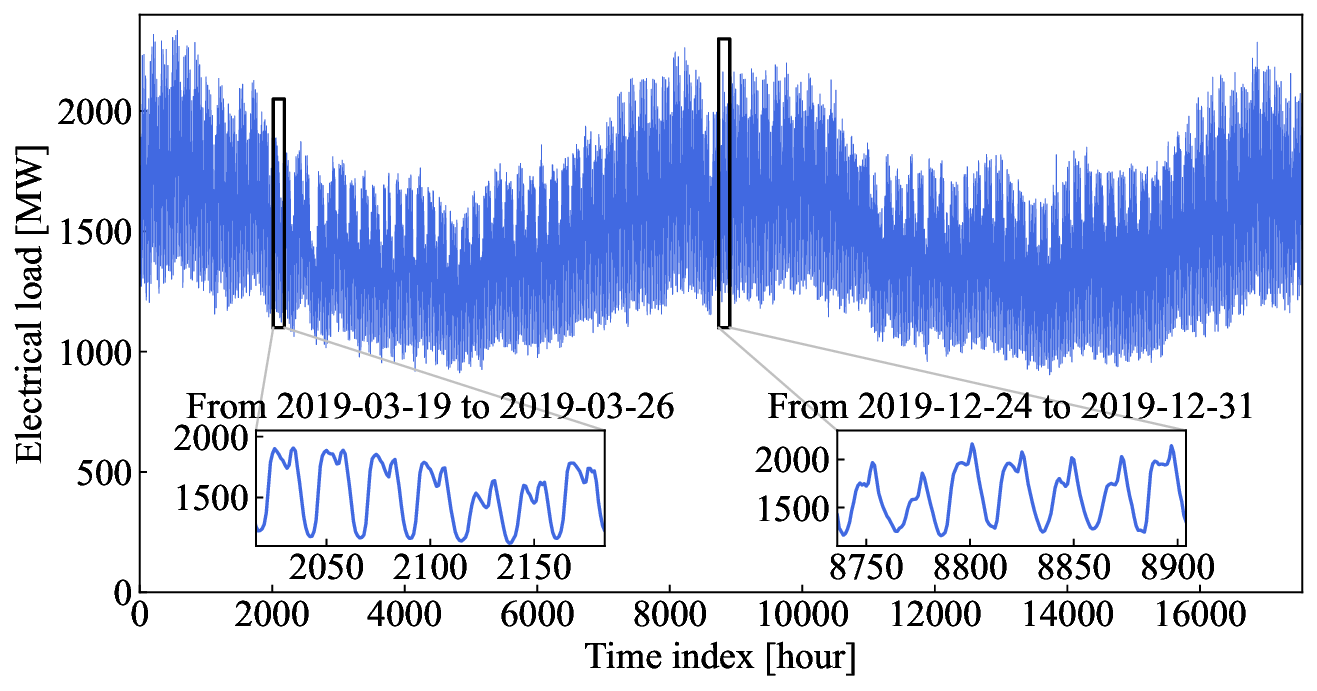}
		\end{minipage}
	}	
	\subfigure[]{
		\begin{minipage} {0.9\linewidth}\label{nor}
			\centering
			\includegraphics[width=1\linewidth]{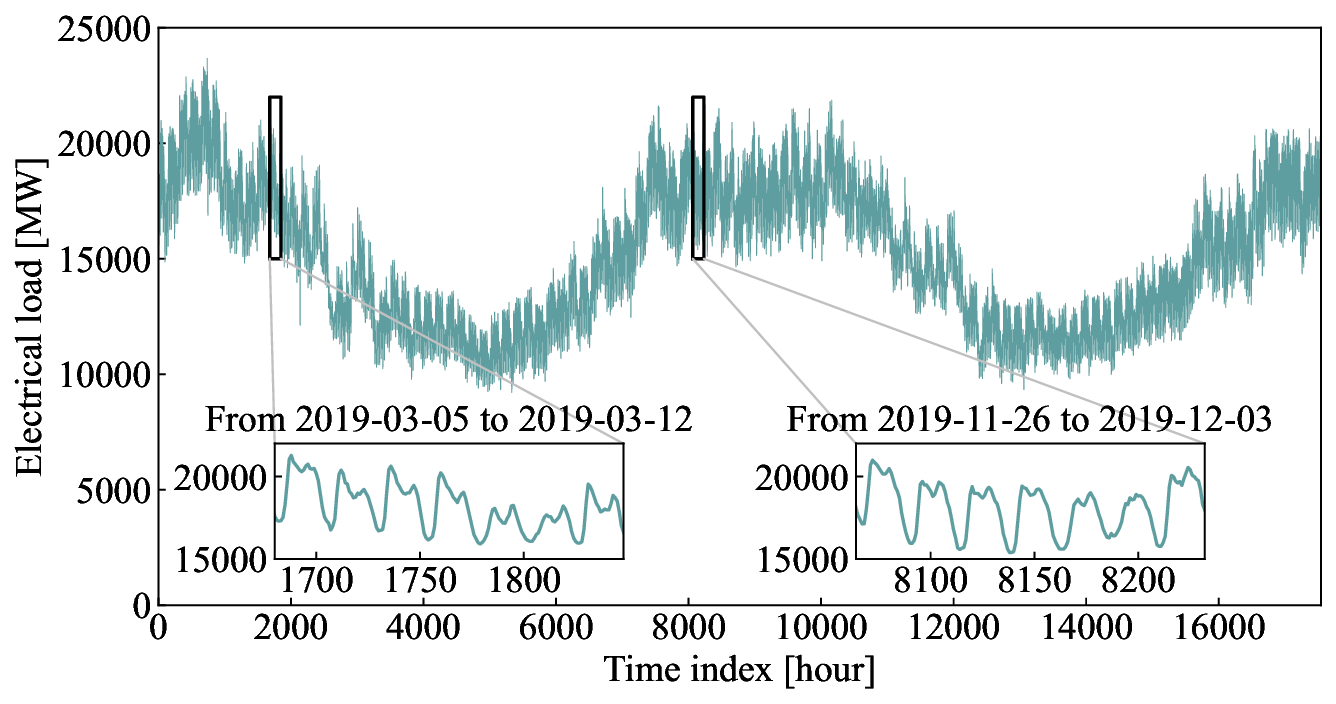}
		\end{minipage}
	}	
	\caption{Electrical load dataset from 2019-01-01 to 2020-12-31 in (a) Belgium, (b) Denmark, and (c) Norway.}
	\label{belgium_load_two_periods}
\end{figure}

Electrical load forecasting plays a fundamental role in the optimal planning and operation of the power system. In this work, three publicly available datasets recording the electrical load of Belgium\cite{elia}, Denmark\cite{energinet}, and Norway \cite{statnett} are selected for the electrical load forecasting algorithm design and validation, each of which contains hourly electrical load data from 2019-01-01 to 2021-12-31.
Here, the electrical load data from 2019-01-01 to 2020-12-31 are utilized as the training dataset, and the rest is the test dataset. The training data for these three countries are illustrated in Fig. \ref{belgium_load_two_periods}, where 2019-01-01 00:00 is set as the first point. Despite being a non-stationary random sequence, the electrical load time series predominantly demonstrates three key characteristics:
\begin{enumerate}
  \item Daily and weekly periodicity. The electrical load in one day experiences a rise, peaking at approximately 9:00 am, and then begins to decline. The weekly electrical load rises on weekdays and decreases on weekends.

  \item Slow trend changes. Some statistical features, such as weekly average load, remain relatively stable in the short term. A noticeable change can only be observed over longer time intervals.

  \item Alike days. There exist alike days in the electrical load time series, i.e., the data and trends on the historical days are similar to those on the forecast days.
\end{enumerate}

\subsection{Feature Selection}

Based on the above analysis, we select features from the following four aspects to build the electrical load forecasting model.

\subsubsection{Historical time series}
The historical sequence of the electrical load data in the past week (168 hours) is utilized to forecast the electrical load on the next day (24 hours). Therefore, a sliding window with a size of 168 data points is adopted and the sliding interval is set as 24, i.e., given the original electrical load dataset $\{l_1, l_2, \cdots, l_N\}$ with $l_j$ ($1 \leq j \leq N$) the electrical load data and $N$ the data number, the $i$-th sliding window is ${\mathbf {L}_i} = [ {l_{24(i-1)+1}},~{l_{24(i-1)+2}},\cdots,$ $~l_{24(i-1)+168}]^T$, and its corresponding forecasted electrical load series is $\mathbf{Y}_i=[{l_{24(i-1)+169}},~{l_{24(i-1)+170}},\cdots,~l_{24(i-1)+192}]^T \in {\mathbb{R}^{24}}$.

\subsubsection{Time index features}
Motivated by the daily and weekly periodicity of electricity load data (as seen in the zoom of Fig. \ref{belgium_load_two_periods}), three features for model construction are extracted from the original dataset $\mathbf{L}$ as follows:

\begin{enumerate}

\item Day index in one week corresponding to the $t$-th time step, denoted as $W_t$, where $W_t=0,1,...,6$ when $t$ represents Monday, Tuesday, $...$, Sunday, respectively.

\item  Time index in one day corresponding to the $t$-th time step, denoted as $T_t$, where $T_t=0,1,...,23$ when $t$ represents 00:00, 01:00, $...$, 23:00, respectively.

\item  Holiday mark for the $t$-th time step, denoted as $H_t$,  where $H_t=0$ or $1$ when the corresponding day is a holiday or not.

\end{enumerate}
Since $W_t$, $T_t$, and $H_t$ are discrete categorical features, it is necessary to transform them into a one-hot encoded form so that they can be better interpreted by the neural network. Referring to \cite{zhi2021con}, the one-hot vectors of $W_t$, denoted as $\tilde{\mathbf{W}}_t \in {\mathbb{R}^7}$,  can be obtained by

\begin{equation} \label{1}
\begin{array}{ll}
\tilde{\mathbf{W}}_t =[\tilde{W}_{t_0}, \cdots, \tilde{W}_{t_i}, \cdots, \tilde{W}_{t_6}]^T \\
\end{array}
\end{equation}
with $\begin{array}{ll}
  \tilde{W}_{t_i}= \left\{ {\begin{array}{rlc}
	1,& \text{if}~{W}_{t} = i \\
	0,&\text{otherwise}
	\end{array}} \right.
\end{array}$.
Similarly, the one--hot vectors of $T_t$ and $H_t$ can be calculated and denoted as $\tilde{\mathbf{T}}_t \in {\mathbb{R}^{24}}$ and $\tilde{\mathbf{H}}_t \in {\mathbb{R}^2}$, respectively.

\subsubsection{Historical statistical features}
\begin{figure}[!htbp]
	\centering		
			\centering
	\includegraphics[width=0.9\linewidth]{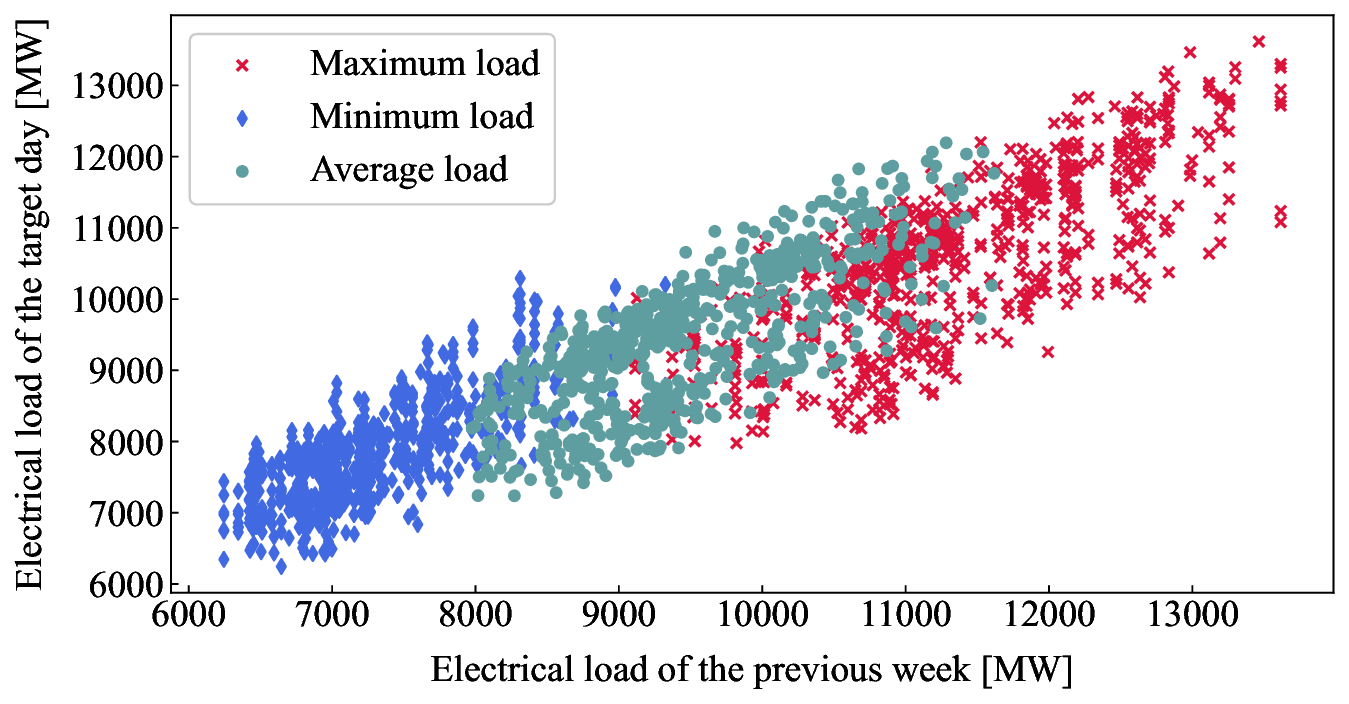}	
	\caption{Correlation of three statistical features between the previous week and target day.}
	\label{statistical features correlation}
\end{figure}

The short-term historical statistical features can reflect the tendency characteristics of the electrical load. Therefore, the following three historical statistical features of the last week before the target day are selected to enhance the forecasting performance: 1) maximum load, 2) minimum load, and 3) average load of the previous week. From Fig. \ref{statistical features correlation}, it is observed a positive correlation between these three features of the previous week and the target day. For the sliding window data $\mathbf{L}_i$, its historical statistical feature vector $\mathbf{F}_{i} \in {\mathbb{R}^{3}}$ can be denoted as:
\begin{equation}
\mathbf{F}_{i} = \left[L_{i, wmax}, L_{i, wmin}, L_{i, wavg} \right]^T
\end{equation}
where $L_{i, wmax}$, $L_{i, wmin}$, and $L_{i, wavg}$ are the maximum, minimum, and average values in $\mathbf{L}_i$, respectively.


\subsubsection{Similarity features}
From Fig. \ref{belgium_load_two_periods}, it is observed that there exist alike days in the electrical load dataset, i.e., the data and trends on the target day are similar to those on some historical days. Hence, we can extract the similarity features between each historical sequence and each repeatedly occurring sequence, and establish the relationship between this similarity and the forecasted electrical load during training to enhance the forecasting accuracy. Here, a K-means-based similarity extraction algorithm \cite{DUDEK2015277} is utilized to identify the historical electrical load time series with similar characteristics and to recognize representative patterns. The detailed process of the similarity extraction algorithm is provided as follows:
\begin{enumerate}
  \item Randomly initialize the set of clustering centers as $\mathcal{C}^{\left( 0 \right)} = \left\{ {{\mathbf{c}_1^{\left( 0 \right)}},{\mathbf{c}_2^{\left( 0 \right)}}, \ldots {\mathbf{c}_{n_c}}^{\left( 0 \right)}} \right\}$, where $\mathbf{c}_i^{\left( 0 \right)} \in \mathbb{R}^{168}$ denotes the $i$-th clustering center and $n_c$ denotes the total number of clusters.

  \item Categorize each electrical load sequence to the nearest clustering centers. The set of clusters at the $k$-th iteration is denoted by $\mathcal{S}^{\left( k \right)} = \left\{ {{\mathcal{S}_1^{\left( k \right)}},{\mathcal{S}_2^{\left( k \right)}}, \ldots {\mathcal{S}_{n_c}}^{\left( k \right)}} \right\}$. For any one sequence $\mathbf{L}_i$ in the training set, search for the ordinal number of the nearest clustering center, which is given by $\mathop {\arg \min }\limits_j \left\| {{\mathbf{L}_i} - \mathbf c_j^{(k)}} \right\|_2$ with $||\cdot||_2$ being the 2-norm, then categorize $\mathbf{L}_i$ to $\mathcal{S}^{(k)}_j$.

  \item Update the clustering centers to minimize the summation of the distance between each sequence and each clustering center. The updating process is given by
    $\mathop {\arg \min }\limits_{\mathcal{C}^{\left( k+1 \right)}} \sum\limits_{j = 1}^{{n_c}} {\sum\limits_{\mathbf L_i \in {S_j}^{\left( k \right)}} {{{\left\| {\mathbf{L}_i - {\mathbf{c}_j^{\left( k+1 \right)}}} \right\|_2}}} }$.
The summation of the distance after the $k$-th updating process can be calculated and denoted as $J(k+1)$.

   \item Terminate and output the optimized set of clustering centers $\mathcal{C} = \mathcal{C}^{\left( k+1 \right)}$, if $|J(k+1)-J(k)| \leq \epsilon$. Otherwise, increase $k$ to $k+1$ and return to Step 2).

  \item The similarity feature vector $\mathbf{P}_i \in  {\mathbb{R}^{n_c}}$ between the sequence $\mathbf{L}_i$ and the historical patterns (i.e., the optimized clustering center $\mathcal{C}$) can be calculated as:
  \begin{equation}
  	{\mathbf{P}_i} = \left[ {{p_{i,1}},{p_{i,2}}, \ldots {p_{i,{n_c}}}} \right]^T
  \end{equation}
with ${p_{i,j}} = \frac{{\mathbf{L}_i^T \cdot \mathbf{c}}_j}{{{{\left\| {\mathbf{L}_i} \right\|_2}}{{\left\| {{\mathbf{c}}_j} \right\|_2}}}}$ denoting the similarity between the $i$-th sequence and $j$-th clustering center.
 
\end{enumerate}
 Note that most of the existing studies related to electrical load forecasting \cite{en11010213, 7748604, 982201, JURADO2015276, kong2017short, BASHIR20221678} only utilize two or three out of the aforegoing four types of features. In this work, by comprehensively incorporating the historical time series, time index features, historical statistical features, and similarity features in the electrical load forecasting model, a more accurate forecast of the electrical load can be expected.

\section{THE FORECASTING MODEL}
The features selected in Section \uppercase\expandafter{\romannumeral2} can be divided into two categories: temporal features (historical time series) and non-temporal features (the rest part in the constructed feature pool). Note that the LSTM neural network is known to perform well for temporal feature-based modeling but the performance could deteriorate when applied to non-temporal features. Conversely, the FCNN demonstrates the opposite behavior. To address these limitations, a hybrid LSTM model structure has previously been proposed and employed in solar irradiance forecasting \cite{HUANG20211041} and electric vehicle charging station occupancy prediction \cite{MA2022123217}, which is composed of an LSTM neural network block to model the temporal feature and a multi-layer FCNN block to model the non-temporal features. Such a model structure can be a good candidate for electrical load forecasting, but this has not been explored before.

\begin{figure}[!htbp]
    \centering
\includegraphics[width=0.9\linewidth]{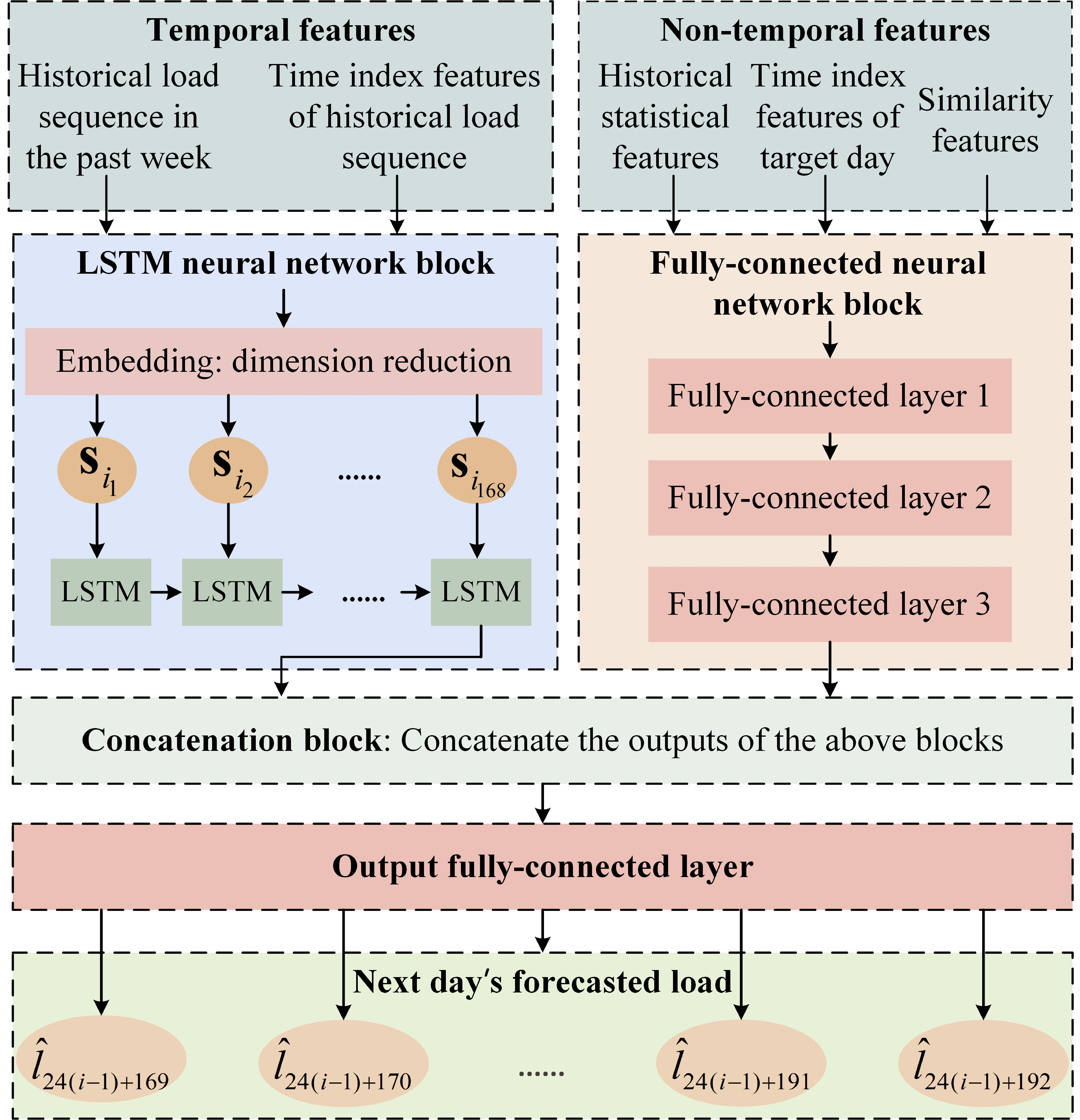}
    \caption{Hybrid LSTM-based electrical load forecasting model scheme.}
    \label{forecasting model}
\end{figure}

Based on this structure, a hybrid LSTM-based model is developed here to forecast the day-ahead electrical load. As illustrated in Fig. \ref{forecasting model}, the forecasting model is composed of an LSTM neural network block to handle the historical electrical load sequence in the last week before the target day $\mathbf{L}_i$ with its time index features, and a multi-layer FCNN block to model the non-temporal features including the time index features of the target day, historical statistical features, and similarity features. Then, a concatenation block is developed to integrate the hidden states obtained by the aforementioned blocks. Finally, an output block consisting of an FC layer is adopted to establish the correlation between the hidden states and the forecasted load.

\subsection{LSTM Neural Network for Temporal Features Modeling}
\begin{figure}[!htbp]
	\centering
	\includegraphics[width=0.9\linewidth]{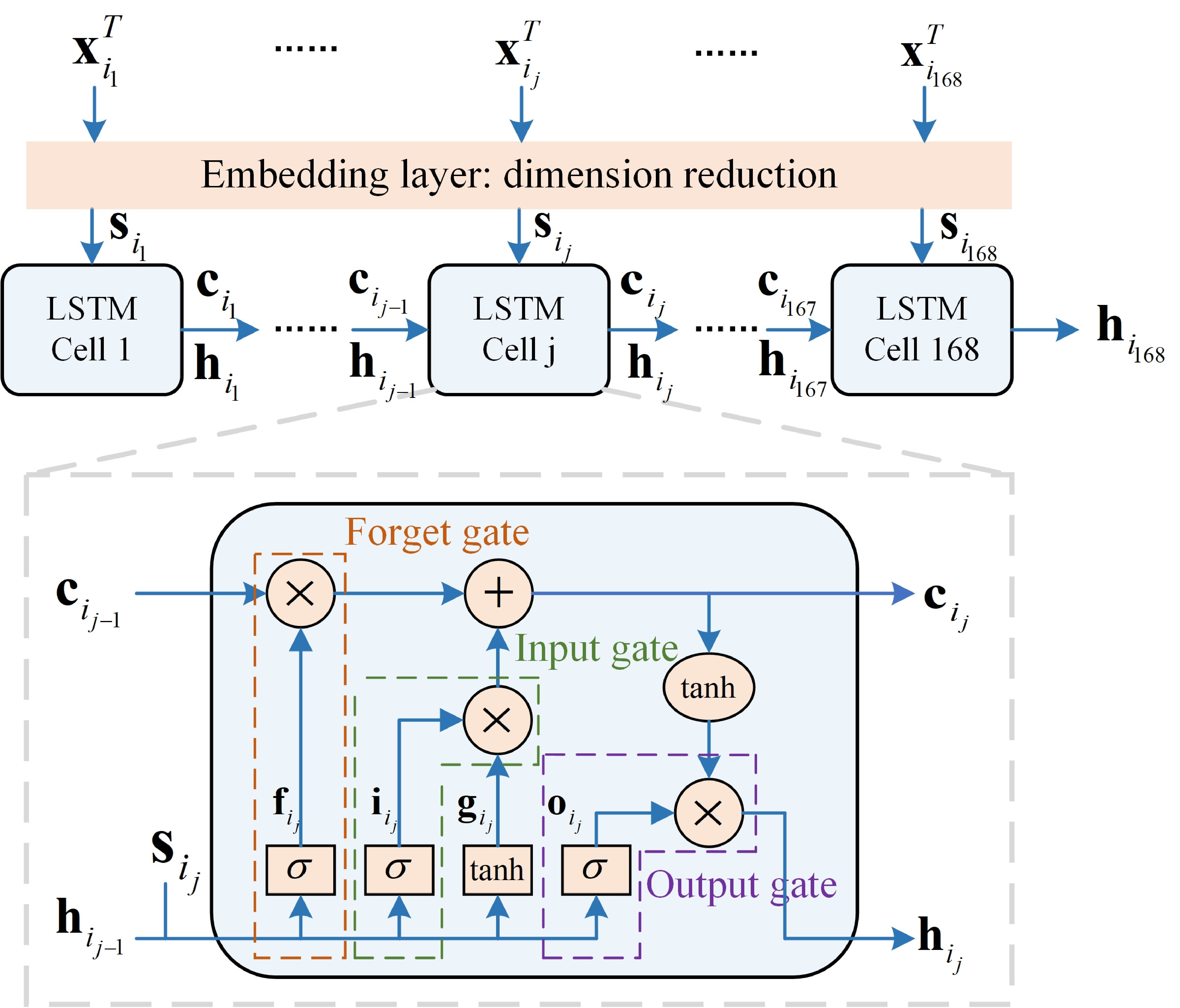}
	\caption{Structure of the LSTM neural network block in the developed model.}
	\label{LSTM}
\end{figure}

The LSTM neural network is first developed in \cite{Neco1997}, which has shown good performance in handling temporal features. The adopted structure of the LSTM neural network block is illustrated in Fig. \ref{LSTM}, and a brief description is presented as follows. For the $i$-th sliding step, the historical time series $\mathbf L_i$ and the time index features corresponding to its elements are utilized as input for the LSTM neural network block, i.e., the input is $\mathbf{X}_i=[\mathbf{x}_{i_1}^T, \mathbf{x}_{i_2}^T, \cdots, \mathbf{x}_{i_{168}}^T]^T \in {\mathbb{R}^{168 \times 34}}$ with $\mathbf{x}_{i_j}=[l_{24(i-1)+j},~ \tilde{\mathbf{W}}_{24(i-1)+j}^T,~  \tilde{\mathbf{T}}_{24(i-1)+j}^T,~  \tilde{\mathbf{H}}_{24(i-1)+j}^T]$.

 Since $\mathbf{x}_{i_j}$ is a high-dimensional and sparse vector because of one-hot encoding, an embedding layer is adopted to transform it into a dense vector, which can significantly reduce the complexity. The $j$-th embedded feature vector $\mathbf{s}_{i_j} \in {\mathbb{R}^{d}}$ ($1 \leq j \leq 168$) for  $i$-th sliding step can be calculated by \cite{PHAM20181}
\begin{equation}
	\begin{array}{ll}
	\mathbf{s}_{i_j} = {\mathbf{W}}_{e}\mathbf{x}_{i_j}^T + \mathbf{b}_{e}
	\end{array}
\end{equation}
where $\mathbf{W}_{e} \in {\mathbb{R}^{d \times 34}} $ and ${\mathbf{b}}_{e} \in {\mathbb{R}^{d}}$ denote the embedding weights and bias, respectively, and $d$ is selected much smaller than $34$ for dimension reduction.

Next, these embedded feature vectors are fed into the LSTM neural network layer for hidden knowledge acquisition, where each LSTM cell takes one embedded feature vector as its input.
As shown in Fig. \ref{LSTM}, the $j$-th LSTM cell mainly consists of three gates, namely the input, forget, and output gates. The input gate $\mathbf{i}_{i_j}$ determines whether the new state information $\mathbf{g}_{i_j}$ can be received, the forget gate $\mathbf{f}_{i_j}$ is responsible for remembering or forgetting the cell state $\mathbf{c}_{i_{j-1}}$ outputted by the previous LSTM cell indexed by ${j-1}$, and the output gate $\mathbf{o}_{i_j}$ decides which information should be outputted as $\mathbf{h}_{i_j}$, which can be calculated as:
\begin{equation} \label{lstm1}
		\begin{array}{ll}
		\mathbf{g}_{i_j}=\tanh(\mathbf{W}_{gh} \cdot\mathbf{h}_{i_{j-1}} + \mathbf{W}_{gs} \cdot \mathbf{s}_{i_j}+\mathbf{b}_{g}) \\[2pt]
		\mathbf{i}_{i_j}=\sigma(\mathbf{W}_{ih} \cdot\mathbf{h}_{i_{j-1}} + \mathbf{W}_{is} \cdot  \mathbf{s}_{i_j}+\mathbf{b}_{i})\\[2pt]	
        \mathbf{f}_{i_j}=\sigma(\mathbf{W}_{fh} \cdot\mathbf{h}_{i_{j-1}} + \mathbf{W}_{fs}  \cdot  \mathbf{s}_{i_j}+\mathbf{b}_{f}) \\[2pt]
			\mathbf{c}_{i_j}=\mathbf{f}_{i_j} \odot \mathbf{c}_{i_{j-1}}+\mathbf{i}_{i_j} \odot \mathbf{g}_{i_j} \\ [2pt]
        \mathbf{o}_{i_j}=\sigma(\mathbf{W}_{oh} \cdot\mathbf{h}_{i_{j-1}} + \mathbf{W}_{os} \cdot  \mathbf{s}_{i_j}+\mathbf{b}_{o}) \\[2pt]
			\mathbf{h}_{i_j}=\mathbf{o}_{i_j} \odot \tanh (\mathbf{c}_{i_j})
		\end{array}
\end{equation}
where $\tanh(\cdot)$ is the activation function of hyperbolic tangent, $\sigma(\cdot)$ is the activation function of sigmoid, $\odot$ represents the element-wise product, $\mathbf{W}_{gh}  \in \mathbb{R}^{n_h \times n_h}$, $\mathbf{W}_{gs} \in \mathbb{R}^{n_h \times d}$, $\mathbf{W}_{ih}  \in \mathbb{R}^{n_h \times n_h}$, $\mathbf{W}_{is}  \in \mathbb{R}^{n_h \times d}$, $\mathbf{W}_{fh} \in \mathbb{R}^{n_h \times n_h}$ $\mathbf{W}_{fs} \in \mathbb{R}^{n_h \times d}$, $\mathbf{W}_{oh} \in \mathbb{R}^{n_h \times n_h}$ and $\mathbf{W}_{os} \in \mathbb{R}^{n_h \times d}$ denote the weight matrices with $n_h$ being the number of LSTM neural network hidden nodes, $\mathbf{b}_{g}  \in \mathbb{R}^{n_h}$, $\mathbf{b}_{i}  \in \mathbb{R}^{n_h}$, $\mathbf{b}_{f} \in \mathbb{R}^{n_h}$, and $\mathbf{b}_{o} \in \mathbb{R}^{n_h}$ are the bias vectors, respectively.


Note that the ability of the LSTM that can selectively store, forget, and output information allows it to capture long-term dependencies in sequential data, thus improving the accuracy of the electrical load. The output of the LSTM neural network block is the output state of the last LSTM cell $\mathbf{h}_{i_{168}}$.

\subsection{FCNN for Non-Temporal Features Modeling}

The LSTM neural network performs poorly when applied to non-temporal data because these data usually lack time dependence.
Hence, a multi-layer FCNN block is developed to model the non-temporal features. For the $i$-th sliding step, the historical statistical features $\mathbf{F}_{i}$, the time index features of target day $\mathbf{\tilde{W}}_{24(i-1)+169}$ and $\mathbf{\tilde{H}}_{24(i-1)+169}$, and the similarity feature vector $\mathbf{P}_i$ are utilized as the input of the FCNN block, which are denoted by $\mathbf{Q}_i = [\mathbf{F}_{i}^T, \mathbf{\tilde{W}}^T_{24(i-1)+169}, \mathbf{\tilde{H}}^T_{24(i-1)+169}, \mathbf{P}_i^T]^T \in \mathbb{R}^{(12+n_c)}$.

One input layer and three FC layers are utilized in the neural network block, where each neuron in the FCNN receives inputs from all neurons in the previous layer, and performs a linear combination and a nonlinear transformation of the inputs based on their corresponding weights and biases, and then passes the results to the next layer. Through multiple iterations of training, the FCNN can learn the complicated relationship from the selected non-temporal features to the next day's electrical load.

Referring to \cite{schwing2015fully}, the output of the FCNN block, denoted by $\mathbf{h}_{i,Q} \in \mathbb{R}^{n_h}$, is calculated by 
\begin{equation}
	{\mathbf{h}_{i,Q}} = {\mathbf{W}}_{f_3}\mathbf{\varphi}\left( {{\mathbf{W}}_{f_2}\mathbf{\varphi}({\mathbf{W}}_{f_1}{{\mathbf{Q}}_i} + {\mathbf{b}}_{f_1}) + {\mathbf{b}}_{f_2}} \right)+ {\mathbf{b}}_{f_3}
\end{equation}
where ${\mathbf{W}}_{f_1} \in \mathbb{R}^{n_h \times (12+n_c)}$, ${\mathbf{W}}_{f_2} \in \mathbb{R}^{n_h \times n_h}$, and ${\mathbf{W}}_{f_3}\in \mathbb{R}^{n_h \times n_h}$ denote the weight matrices,  ${\mathbf{b}}_{f_1} \in \mathbb{R}^{n_h}$, ${\mathbf{b}}_{f_2} \in \mathbb{R}^{n_h}$, and ${\mathbf{b}}_{f_3} \in \mathbb{R}^{n_h}$ are the bias vectors, and $\varphi(\cdot)$ is the activation function of ReLU, which is defined as   $\varphi (z) = \max (z,0)$.

\subsection{Block Concatenation and Output}
After modeling the temporal and non-temporal features by the LSTM and FCNN blocks, respectively, it is essential to integrate their outputs into a single-vector representation. To achieve this, a concatenation block is developed by $\mathbf{h}_{i,f}=[\mathbf{h}_{i_{168}}^T,\mathbf{h}_{i,Q}^T]^T$, where $\mathbf{h}_{i,f}\in {\mathbb{R}^{{2n_h}}}$ is the concatenated vector and then used as the input to the output layer in the designed neural network model to obtain the next day's forecasted electrical load:
\begin{equation}
    {{{\mathbf{\hat Y}}}_i} = {\mathbf{W}}_{o_2}\mathbf{\varphi}({\mathbf{W}}_{o_1}{\mathbf{h}_{i,f}} + {\mathbf{b}}_{o_1}) + {\mathbf{b}}_{o_2}
\end{equation}
where ${\mathbf{W}}_{o_1} \in \mathbb{R}^{n_h \times 2n_h}$ and ${\mathbf{W}}_{o_2} \in \mathbb{R}^{24 \times n_h}$ denote the weight matrices, and ${\mathbf{b}}_{o_1} \in \mathbb{R}^{n_h}$ and ${\mathbf{b}}_{o_2} \in \mathbb{R}^{24}$ are the bias vectors, respectively.

\section{MODEL TRAINING AND ONLINE CORRECTION} 
To obtain accurate forecasting results of the day-ahead electrical load, the parameters of the developed hybrid LSTM-based model should be derived based on the training dataset. Moreover, the online correction strategy is designed based on the latest electrical load sequences to further improve the forecasting performance. 

\subsection{Offline Model Training based on Gradient Regularization}
During the offline training phase, the following two objectives should be considered: 1) making forecasting results approximate to the true values in the training set to a certain extent, and  2) ensuring the anti-disturbance capability of the forecasting model. Inspired by \cite{ross2018improving}, the input gradient regularization method is applied here to generate adversarial samples for the anti-disturbance capability training of the forecasting model. However, considering the input features contain one-hot vectors which do not admit infinitesimal perturbation, we turn to perturb the embedding layer in a manner that increases the local linear approximation of the loss function \cite{miyato2016adversarial}. The detailed process to train the electrical load forecasting model offline is given as follows:
\begin{enumerate}
	\item Randomly initialize the set of the forecasting model parameters as $\mathbf{W}^{(0)}$, which is comprised of $\mathbf{W}_e^{(0)}$, denoting the set of the embedding layer parameters, and $\mathbf{W}_r^{(0)}$, denoting the set of the rest parameters.  Initialize the ordinal number of iterations $k=0$, the early stopping patience $j=0$, and the early stopping tolerance $j_{max}$.
	
	\item In the $k$-th iteration, randomly select $N_t$ samples from the training set as a batch $\mathcal{B}_k=\{\left(\{{\mathbf X_i},{\mathbf Q_i}\},{\mathbf Y_i} \right)|i = $ $1,2, \ldots,{N_t} \}$, then calculate the approximation loss to measure the distance between forecasting results and true values, using the following formulation:
	\begin{equation}\label{gamma 122}
		{\Gamma _1}\left( {{\mathcal{B}_k},\mathbf{W}^{(k)}} \right) = \frac{1}{{{N_t}}}\sum\limits_{i = 1}^{{N_t}} {{{\left\| {{{\mathbf{Y}}_i} - {{{\mathbf{\hat Y}}}_i}} \right\|}_1}}
	\end{equation}
	where ${\Gamma _1}(\cdot)$ is the approximation loss function,  ${{\mathbf{\hat Y}}}_i$ is the estimation of $\mathbf{Y}_i$ obtained by the designed forecasting model with the parameters $\mathbf{W}^{(k)}$, and $\left\| \cdot \right\|_1$ denotes the 1-norm.

	\item Calculate the approximation loss function in the presence of embedding layer perturbations \cite{ross2018improving} that
\begin{equation}
	\begin{array}{ll} \label{333}
	{\Gamma}\left( {{\mathcal{B}_k}, \mathbf W^{(k)}} \right) = {\Gamma _1}\left( {{\mathcal{B}_k}, \{{\mathbf W_e^{(k)}} + \Delta {\mathbf W_e^{(k)}}, \mathbf W_r^{(k)}\}} \right)
	\end{array}
\end{equation}
where $\Delta {\mathbf W_e^{(k)}} = \lambda {\nabla _{{\mathbf{W}}_{e}^{(k)}}}\left({\Gamma_1} \left( {{\mathcal{B}_k},{{\mathbf{W}^{(k)}}}} \right)\right)$ is the perturbation vector, where $\lambda$ denotes the scale of perturbation, and $\nabla_{{\mathbf{W}}_{e}^{(k)}}(\cdot)$ represents the gradient with respect to the parameters ${\mathbf{W}}_{e}$.	

	\item To defend against perturbations and approach the true value, the loss function is reduced by back-propagation.
	
	 Calculate the gradient of ${\Gamma}\left( {{\mathcal{B}_k}, \mathbf W^{(k)}} \right)$ in (\ref{333}) on $\mathbf{W}$ that
	\begin{equation}
		\begin{array}{ll}
			\Delta \mathbf{W}^{(k)} =  \nabla_{\mathbf{W}^{(k)}} ({\Gamma _1}\left( {{\mathcal B_k},\mathbf W^{(k)}} \right) + \frac{1}{2}\lambda \\
	~~~~~~~~~~~~	\times \left\| {{\nabla _{{\mathbf W_e^{(k)}}}}\left({\Gamma _1}\left( {{\mathcal B_k},\mathbf W^{(k)}} \right)\right)} \right\|^2_2)
      \end{array}
	\end{equation}
	Then, update the model parameters by $\mathbf{W}^{(k+1)}=\mathbf{W}^{(k)}-\eta^{(k)} \Delta \mathbf{W}^{(k)}$, where $\eta^{(k)}$ denotes the update step in the $k$-th iteration, which is automatically calculated by the Adam optimizer \cite{kingma2014adam}.
	A certain percentage of samples in the training set are randomly selected as the validation set $\mathcal{B}_v$.
	Calculate the loss on validation set $\Gamma\left(\mathcal{B}_v,\mathbf{W}^{(k+1)}\right)$. Increase $j$ to $j+1$, if $|\Gamma\left(\mathcal{B}_v,\mathbf{W}^{(k)}\right)-\Gamma\left(\mathcal{B}_v,\mathbf{W}^{(k+1)}\right)| < \epsilon$. Otherwise, set $j=0$.
	
	\item Terminate and output the optimized parameters $\mathbf{W}^*=\mathbf{W}^{(k)}$, if $j=j_{max}$ or $k$ reaches the maximum number of iterations. Otherwise, increase $k$ to $k+1$, and return to Step 2).
	
\end{enumerate}

\subsection{Online Model Correction by Fine-Tuning Output Layer Parameters}
Since the distribution of electrical load data changes with time, the forecasting error of the developed hybrid LSTM-based model may increase over time with the offline optimized parameters $\mathbf{W}^*$. To mitigate this issue, it is important to periodically retrain the model with updated data. This practice ensures that it remains calibrated to the current distribution of the electrical load data to improve the accuracy of forecasting results. Here, an online model correction strategy \cite{li2018learning} is adopted to fine-tune the parameters in the output FC layer block based on the latest data once a week. Compared with retraining all model parameters, this way of model correction may maintain the anti-disturbance capability achieved from offline training, and lower computational costs and higher accuracy.

For the offline trained model parameters $\mathbf{W}^*$, except the parameters in the output FC layer block, denoted as $\mathbf{W}_o^*$, will be updated, while the rest of the parameters, denoted as $\mathbf{W}_f^*$, are frozen.
The model parameters are updated once a week based on the electrical data from the latest three months. For the $k$-th correction, the updated model parameters can be obtained by ${\mathbf{W}}_{c_{k+1}}^*=\{\mathbf{W}_f^*, \mathbf{W}_{o_{k}^*} + \Delta {{\mathbf{W}}_{o_k}} \}$,
where the initial value of $\mathbf{W}_{o_{k}}^*$ is $\mathbf{W}_o^*$. Similar to offline model training, the updated amount  $\Delta {{\mathbf{W}}_{o_{k}}}$ is calculated as:
\begin{equation}
	\mathop {\arg \min }\limits_{\Delta {{\mathbf{W}}_{o_{k}}}} \Gamma \left( {\mathcal{B}_{o_{k}},\left\{{\mathbf{W}}_f^*, {{\mathbf{W}}_{o_{k}}^* + \Delta {{\mathbf{W}}_{o_{k}}}} \right\}} \right)
\end{equation}
where $\mathcal{B}_{o_{k}}$ is the new sampled batch in the $k$-th correction.
\section{EXPERIMENTS AND DISCUSSIONS}

\subsection{Experimental Configurations}
To validate the performance of the designed hybrid LSTM-based electrical load forecasting model, extensive experiments were conducted in a virtual environment with Python 3.8.8, Tensorflow 2.6.0, and Keras 2.6.0, where the model was trained on a single NVIDIA GTX 1080 TI GPU. The detailed experimental configurations are set as shown in TABLE \ref{configuration details}.

\begin{table}[htbp]
\caption{Configuration Details}
\renewcommand\arraystretch{0.9}
\begin{center}
\setlength{\tabcolsep}{3.3mm}{
\begin{tabular}{ll}
\toprule
\multicolumn{1}{c}{\textbf{Parameter}}&\multicolumn{1}{c}{\textbf{Configuration}} \\
\midrule
\multirow{7}*{Dataset} & {- Training set: Data from 2019 to 2020} \\
& {- Test set: Data in 2021} \\
& {- Validation set: 10$\%$ of the training set} \\
& {- Width of sliding window: 168} \\
& {- Forecasted period: Ahead 24 hours} \\
& {- Normalization: min-max} \\
& {- Number of clusters: 20}\\
\midrule
\multirow{5}*{Forecasting model} & {- Number of embedding layer hidden nodes: 10} \\
& {- Number of LSTM neural network hidden}\\
&{\ \  nodes: 128} \\
& {- Number of FCNN block hidden nodes: 128} \\
& {- Number of output FC layer hidden nodes: 128} \\
\midrule
\multirow{5}*{Offline training}
& {- Scale of perturbation: 1} \\
& {- Optimizer: Adam (learning rate: 0.005)} \\
& {- Batch size: 56} \\
& {- Total number of epochs: 150} \\
& {- Early stopping patience: 7} \\

\midrule
\multirow{5}*{Online correction} & {- Frequency: once a week} \\
& {- Optimizer: Adam (learning rate: 0.01)} \\
& {- Batch size: 56} \\
& {- Total number of epochs: 10} \\
& {- Early stopping tolerance: 5} \\
\bottomrule
\end{tabular}}
\end{center}
\label{configuration details}
\end{table}


To measure the quality of forecasting performance, the following three statistical indicators, i.e., mean absolute error (MAE), mean absolute percentage error (MAPE), and root mean squared error (RMSE), are adopted here as:
\begin{equation}
	\begin{array}{ll}
	\text{MAE} = \frac{1}{{N_{2}-N_{1}+1}}
	\sum\limits_{i = N_{1}}^{{N_2}} {| {l_i-{{\hat l}_{i}}}|}\\
        \text{MAPE} = \frac{1}{{N_{2}-N_{1}+1}}\sum\limits_{i = N_{1}}^{{ N_{2}}} {\frac{{| {{l_{i}} - {{\hat l}_{i}}} |}}{{| {{l_{i}}} |}}} \times 100\% \\
        {\text{RMSE}} = \sqrt {\frac{1}{{{N_{2}-N_{1}+1}}}\sum\limits_{i = N_{1}}^{{ N_{2}}} {{{( {{l_{i}} - {{\hat l}_{i}}} )}^2}} }
        \end{array}
\end{equation}
where $N_1$ and $N_2$ are the indexes of the first and last forecasted data, $l_i$ and $\hat{l}_i$ denote the actual and forecasted electrical loads, respectively.

\subsection{Experimental Results and Sensitivity Analysis}
\subsubsection{Experimental results}
\begin{figure}[!htbp]
	\centering		
	\subfigure[]{
		\begin{minipage} {0.9\linewidth}\label{bel}
			\centering
			\includegraphics[width=1\linewidth, clip]{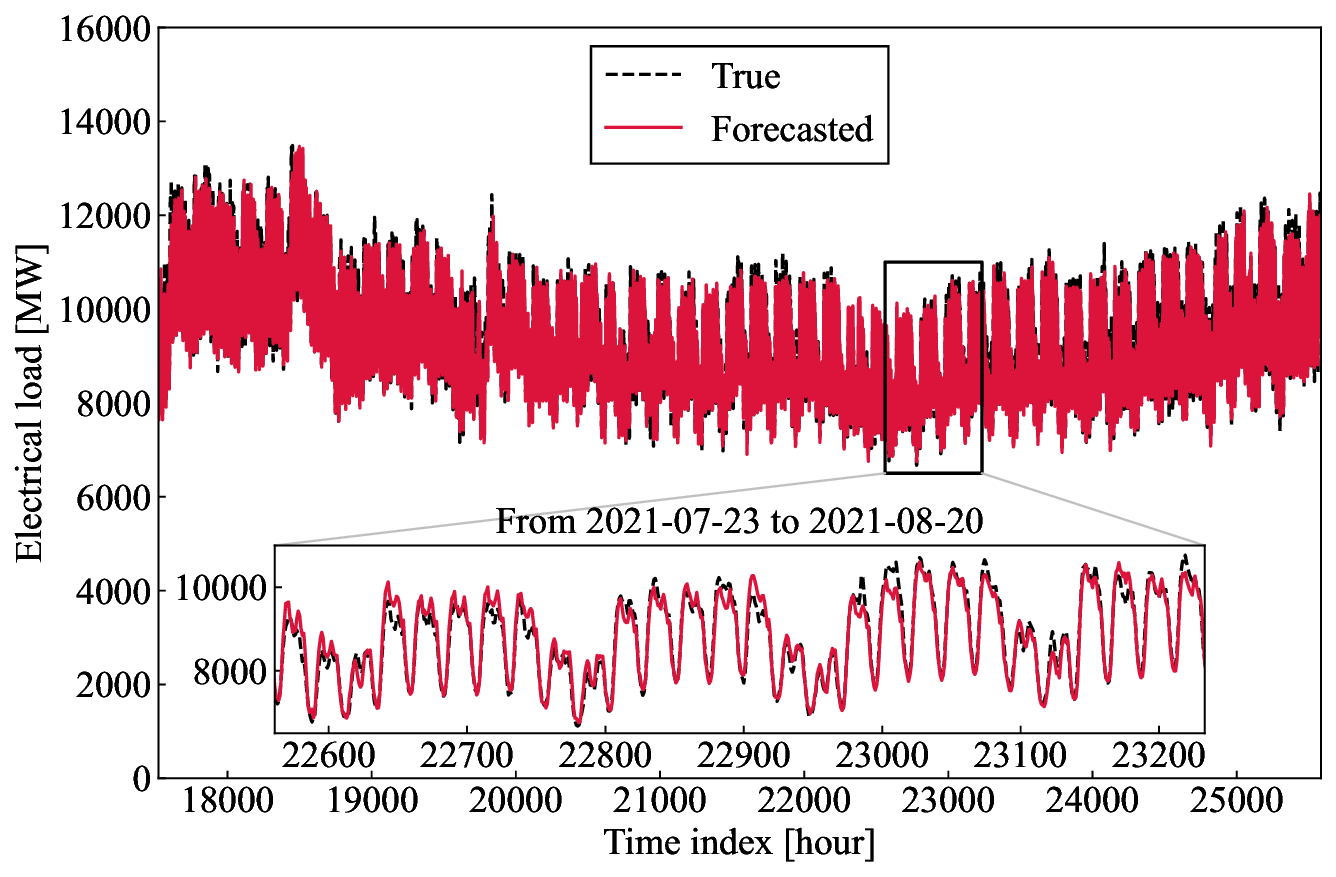}
		\end{minipage}
	}	
	\subfigure[]{
		\begin{minipage} {0.9\linewidth}\label{den}
			\centering
			\includegraphics[width=1\linewidth]{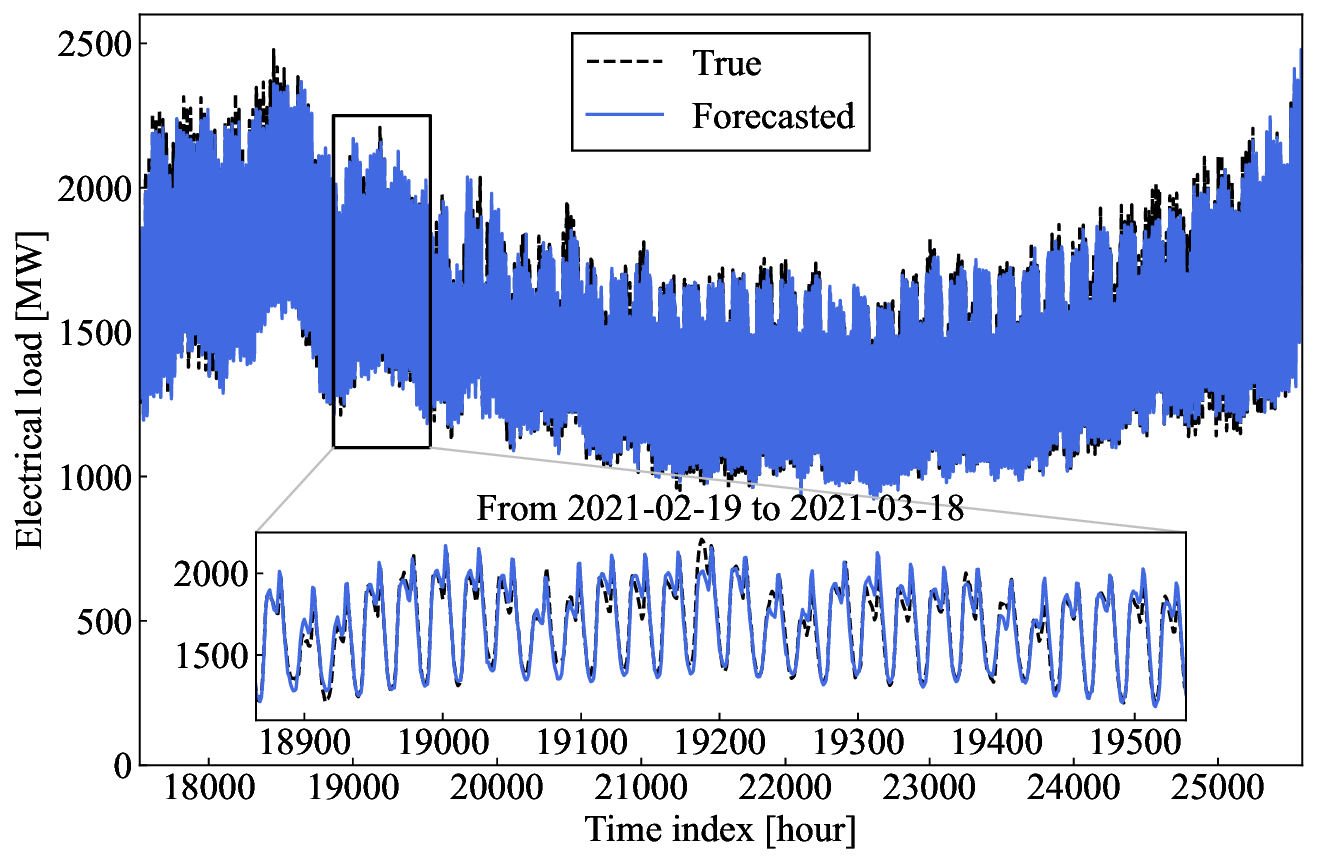}
		\end{minipage}
	}	
	\subfigure[]{
		\begin{minipage} {0.9\linewidth}\label{nor}
			\centering
			\includegraphics[width=1\linewidth]{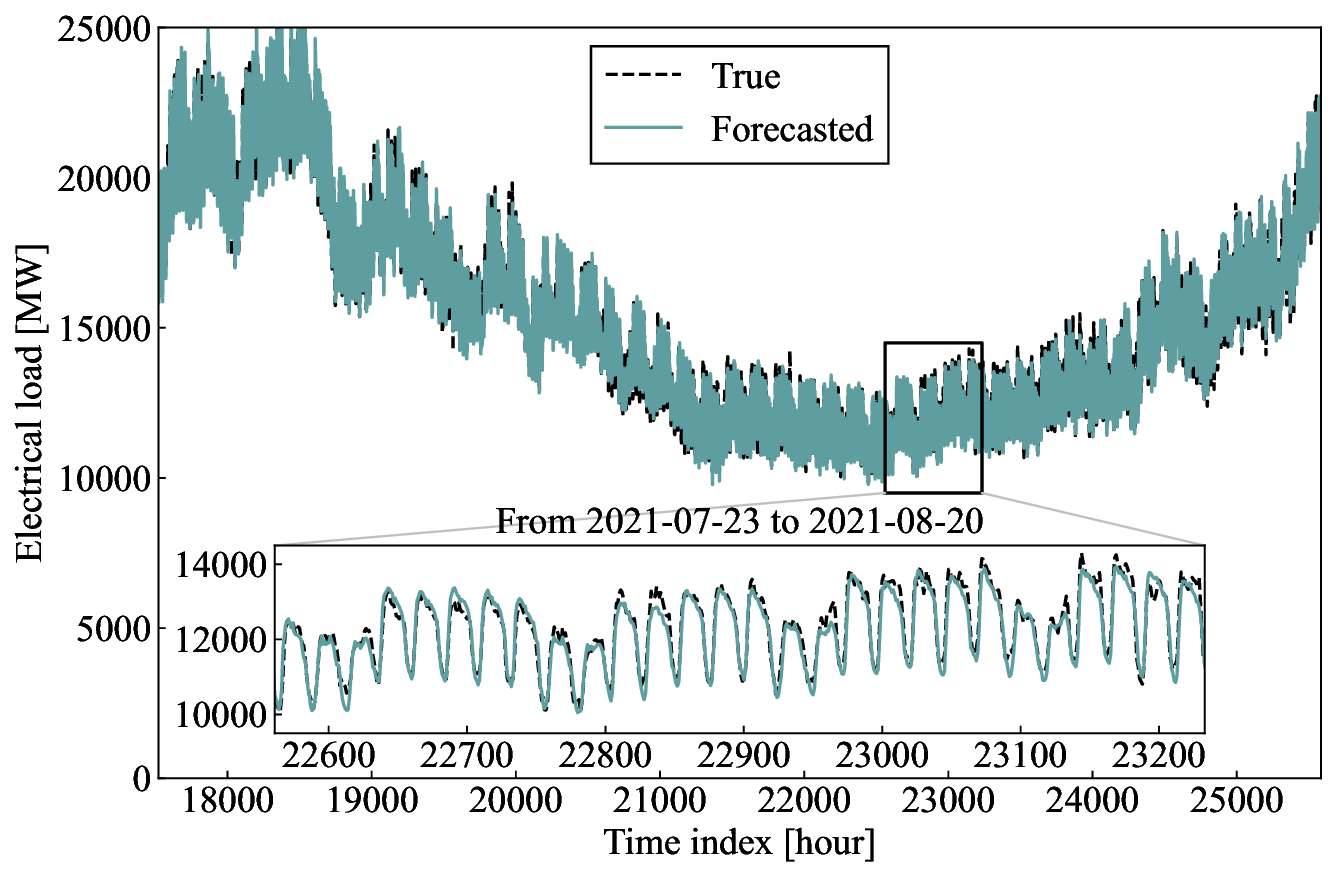}
		\end{minipage}
	}	
	\caption{The forecasting results of the day-ahead electrical load by the developed model from 2021-01-01 to 2021-12-31 in (a) Belgium, (b) Denmark, and (c) Norway.}
	\label{forecasting results}
\end{figure}

The experimental results in terms of the day-ahead forecasted electrical load in 2021 for Belgium, Denmark, and Norway using the developed model are illustrated in Fig. \ref{forecasting results}, in which the zooms of three four-week periods are also given to show details more clearly. It is observed that the forecasted electrical load is highly accurate during valley hours (daily from 0:00 to 11:59 and from 20:00 to 23:59). Despite some small-scale deviations in the forecasted load during peak hours (daily from 12:00 to 19:59), the developed model is able to provide reliable forecasts for most of the studied days. Specifically, the statistical indicators of the forecasting errors in 2021 are provided in TABLE \ref{feature selection}, where the MAE of the model is $255.718\ \text{MW}$/$42.550\ \text{MW}$/$339.112\ \text{MW}$, the MAPE is $2.640\%$/$2.692\%$/$2.152\%$, and the RMSE is 352.044\ MW/ 59.910\ MW/457.098\ MW for Belgium/Denmark/Norway.

\subsubsection{Sensitivity analysis}
\begin{figure}[!htbp]
	\centering		
			\subfigure[RMSE in Belgium]{
			\begin{minipage} {0.45\linewidth}
				\centering
				\includegraphics[width=1\linewidth, clip]{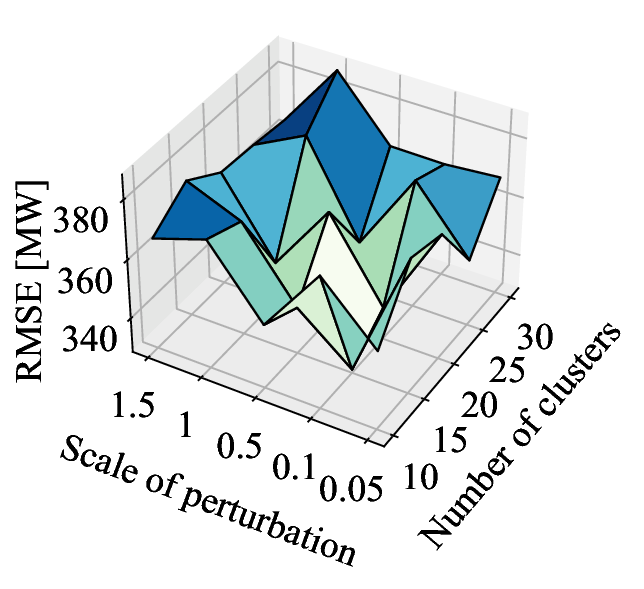}
			\end{minipage}
		}	
			\subfigure[MAPE in Belgium]{
			\begin{minipage} {0.45\linewidth}
				\centering
				\includegraphics[width=1\linewidth]{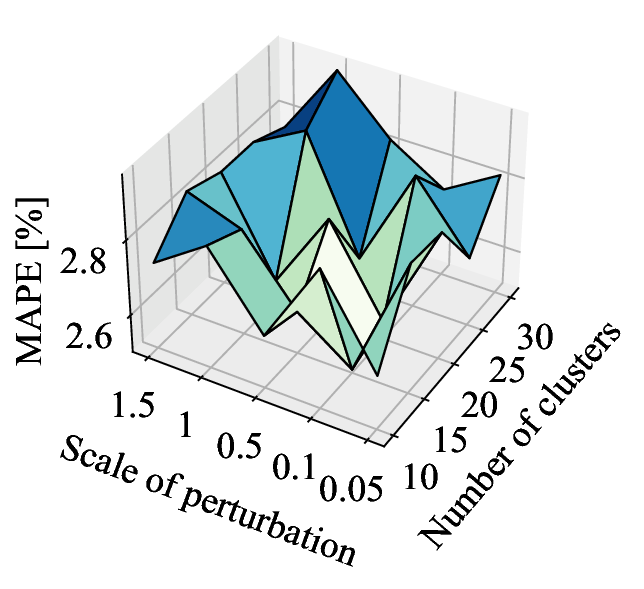}
			\end{minipage}
		}	
		\subfigure[RMSE in Denmark]{
		\begin{minipage} {0.45\linewidth}
			\centering
			\includegraphics[width=1\linewidth]{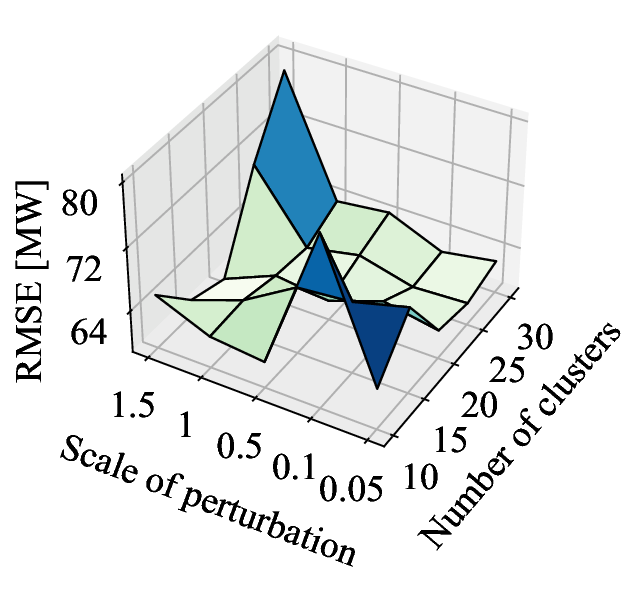}
		\end{minipage}
	}
    \subfigure[MAPE in Denmark]{
			\begin{minipage} {0.45\linewidth}
				\centering
				\includegraphics[width=1\linewidth, clip]{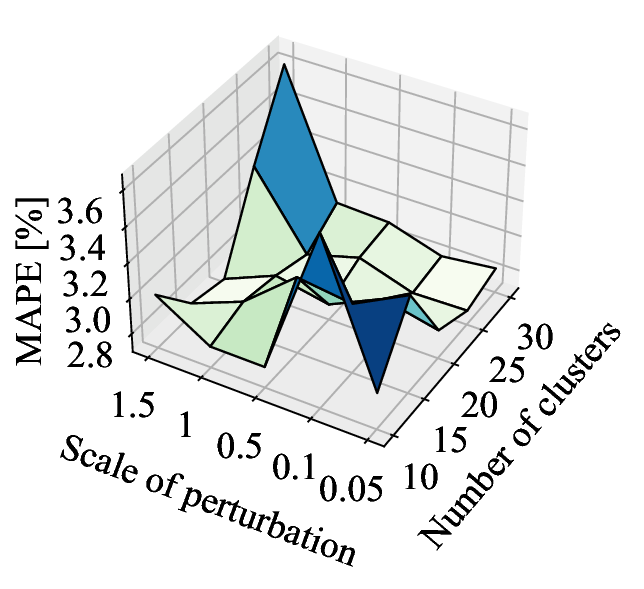}
			\end{minipage}
		}	
			\subfigure[RMSE in Norway]{
			\begin{minipage} {0.45\linewidth}
				\centering
				\includegraphics[width=1\linewidth]{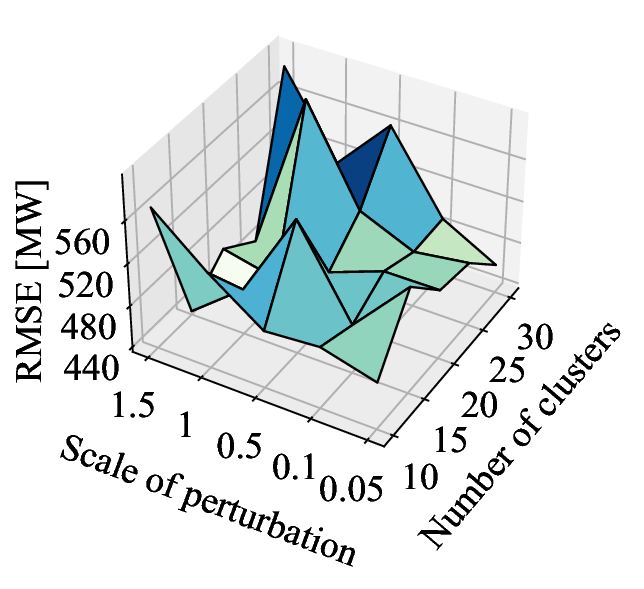}
			\end{minipage}
		}	
		\subfigure[MAPE in Norway]{
		\begin{minipage} {0.45\linewidth}
			\centering
			\includegraphics[width=1\linewidth]{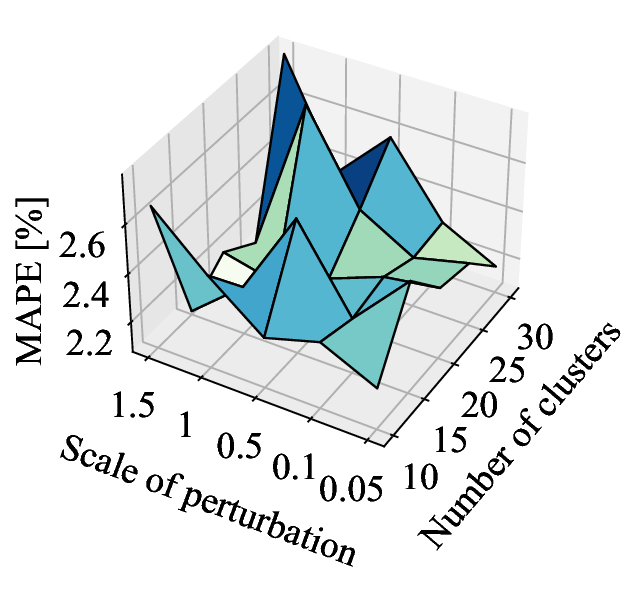}
		\end{minipage}
	}	
	\caption{Results of RMSE and MAPE with different hyperparameter settings.}
	\label{sensitivity analysis}
\end{figure}

In the designed electrical load forecasting strategy, two hyperparameters are related to the forecasting accuracy, i.e., the number of clusters $n_c$ and the scale of perturbation $\lambda$. To evaluate the impacts of different hyperparameter settings on the forecasting results,  $n_c$ is selected as $10$, $15$, $20$, $25$, and $30$, and $\lambda$ is selected as $0.05$, $0.1$, $0.5$, $1$, and $1.5$, sequentially. The corresponding RMSEs and MAPEs of the forecasting results are illustrated in Fig. \ref{sensitivity analysis}, where the model with the hyperparameters $n_c$ of 20 and $\lambda$ of 1 shows excellent forecasting performance for all the three considered countries. Hence, these hyperparameters are selected in the designed electrical load forecasting model in this study.

\begin{center}
	\begin{table}[!htb]
		\renewcommand\arraystretch{0.9}
		\caption{Day-ahead forecasting results in 2021 of the designed model with different feature selections}
		\begin{center}
			\setlength{\tabcolsep}{2mm}{
				\begin{tabular}{llccc}
					\toprule
					\textbf{Dataset} &\textbf{Method} & \textbf{{MAE} [MW]}& \textbf{{MAPE} [\%]}& \textbf{{RMSE} [MW]}\\
					\midrule
					\multirow {4}*{Belgium}&{Proposed}& \textbf{255.718} & \textbf{2.640} & \textbf{352.044}\\
                    &Model 1& 434.464 & 4.483 & 571.784\\
					&Model 2& 272.549 & 2.806 & 368.454\\
					&Model 3& 281.768 & 2.901 & 383.301\\
					\midrule
                    \multirow {4}*{Denmark}&{Proposed}& \textbf{42.550} & \textbf{2.692} & \textbf{59.910} \\
                    &Model 1& 78.256 & 4.720 & 116.494 \\
					&Model 2& 52.866 & 3.378 & 72.415 \\
					&Model 3& 51.003 & 3.260 & 71.595 \\
                    \midrule
                    \multirow {4}*{Norway}&{Proposed}& \textbf{339.112} & \textbf{2.152} & \textbf{457.098} \\
                    &Model 1& 492.749 & 3.141 & 644.967 \\
					&Model 2& 419.383 & 2.645 & 559.874 \\
					&Model 3& 423.527 & 2.660 & 575.701 \\
                    \bottomrule
			\end{tabular}}
		\end{center}
		\label{feature selection}
	\end{table}
\end{center}

\subsection{Comparison with Models Derived From Part of Feature Set}
\begin{figure}[!htbp]
  \centering
  \includegraphics[width=0.9\linewidth]{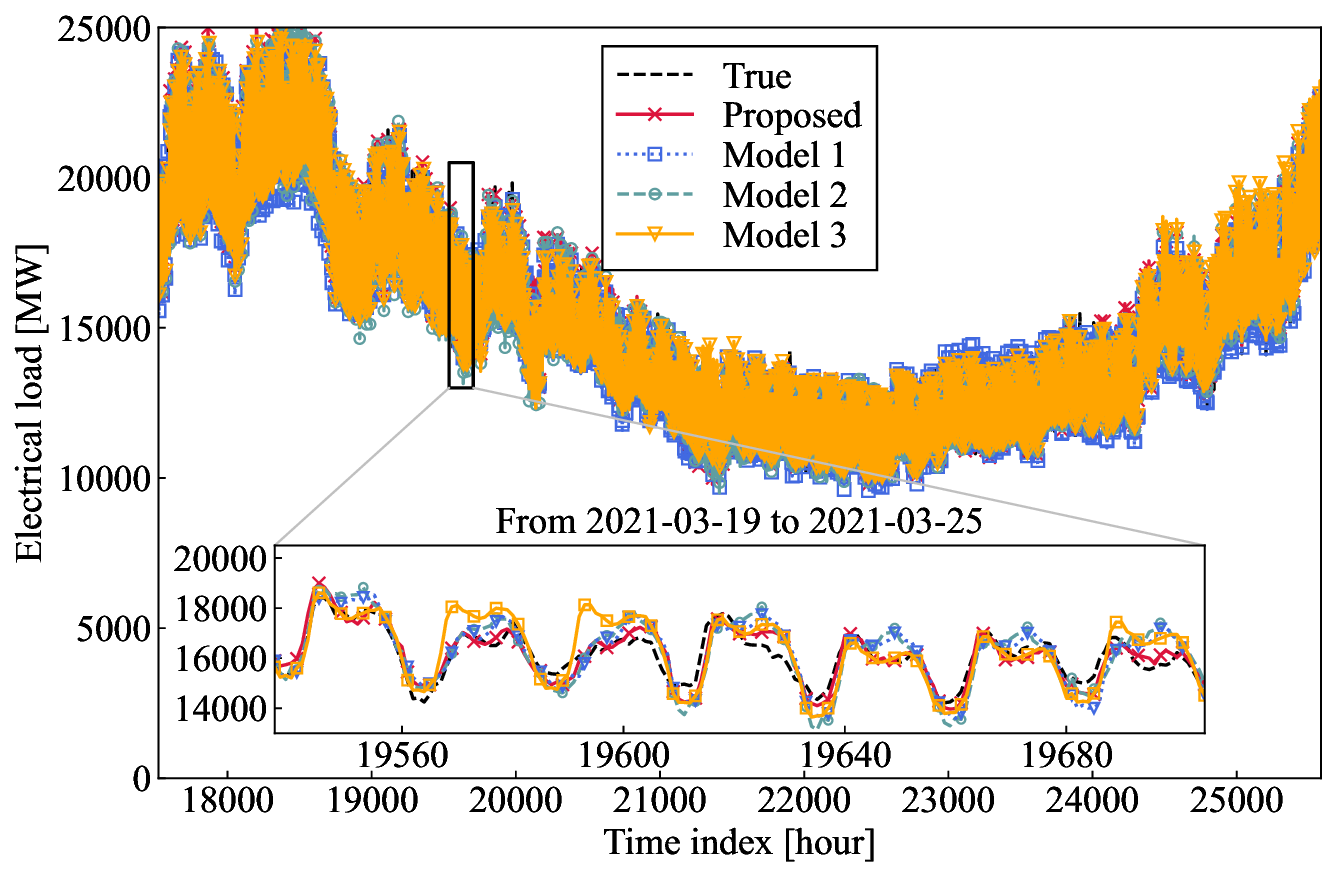}
  \caption{Forecasting results of the designed model with different features in Norway.}
  \label{feature comparison norway}
\end{figure}

To examine the forecasting performance due to the selection of comprehensive and appropriate features in this work, the forecasting results for Norway using different parts of the feature set are provided here for comparison, which are
\begin{enumerate}
\item Model 1: The model only with historical time series.
\item Model 2: The model with historical time series, time index features, and historical statistical features.
\item Model 3: The model with historical time series, time index features, and similarity features.
\end{enumerate}
By using Norway as an example, the day-ahead electrical load forecasting results of the designed model with different feature selections are shown in Fig. \ref{feature comparison norway}. It is observed that more accurate forecasting results can be obtained by comprehensively considering all the four types of features in this work compared to the models only with part of the features. In more detail, Model 3 can give almost comparable results to the proposed model at some times, but obvious deviations can be seen in peak hours on some days (e.g., March 20, 21, and 25).
Compared to the proposed model, significant forecasting errors under Models 1 and 2 can be seen on some target days, such as March 19 and 23-25, as shown in Fig. \ref{feature comparison norway}. 

To quantify the negative impacts of the removal of any type of features on forecasting accuracy, the detailed statistical results of the designed forecasting model with different feature selections are provided in TABLE \ref{feature selection}.
It can be seen that ignoring the similarity features increases the MAE by 6.58\%/24.24\%/23.67\%, and ignoring the statistical features can increase the MAE by 10.19\%/19.87\%/24.89\% in Belgium/Denmark/Norway. Model 1, only considering the temporal feature of historical time series, performs the worst on all the three datasets. The results imply the superior performance offered by modeling all four types of features here, where the time index features are utilized to model periodicity, the statistical features are used for trending change modeling, and the similarity features are adopted to model the relationship with historical alike days.

\subsection{Comparison with State-of-the-Art Forecasting Methods}
To further evaluate the efficacies of the developed model, several state-of-art models for electrical load forecasting are set as benchmarks in this study, including  LSTM of \cite{kong2017short}, convolutional neural network-LSTM (CNN-LSTM) developed in \cite{kim2019predicting}, temporal-attention LSTM (TA-LSTM) of \cite{zang2021residential}, time-dependent CNN (TD-CNN) of \cite{han2018enhanced}, temporal convolutional network (TCN) of \cite{zhu2020short} and FCNN of \cite{ryu2016deep}.

\begin{figure}[!htbp]
  \centering
  \includegraphics[width=0.9\linewidth]{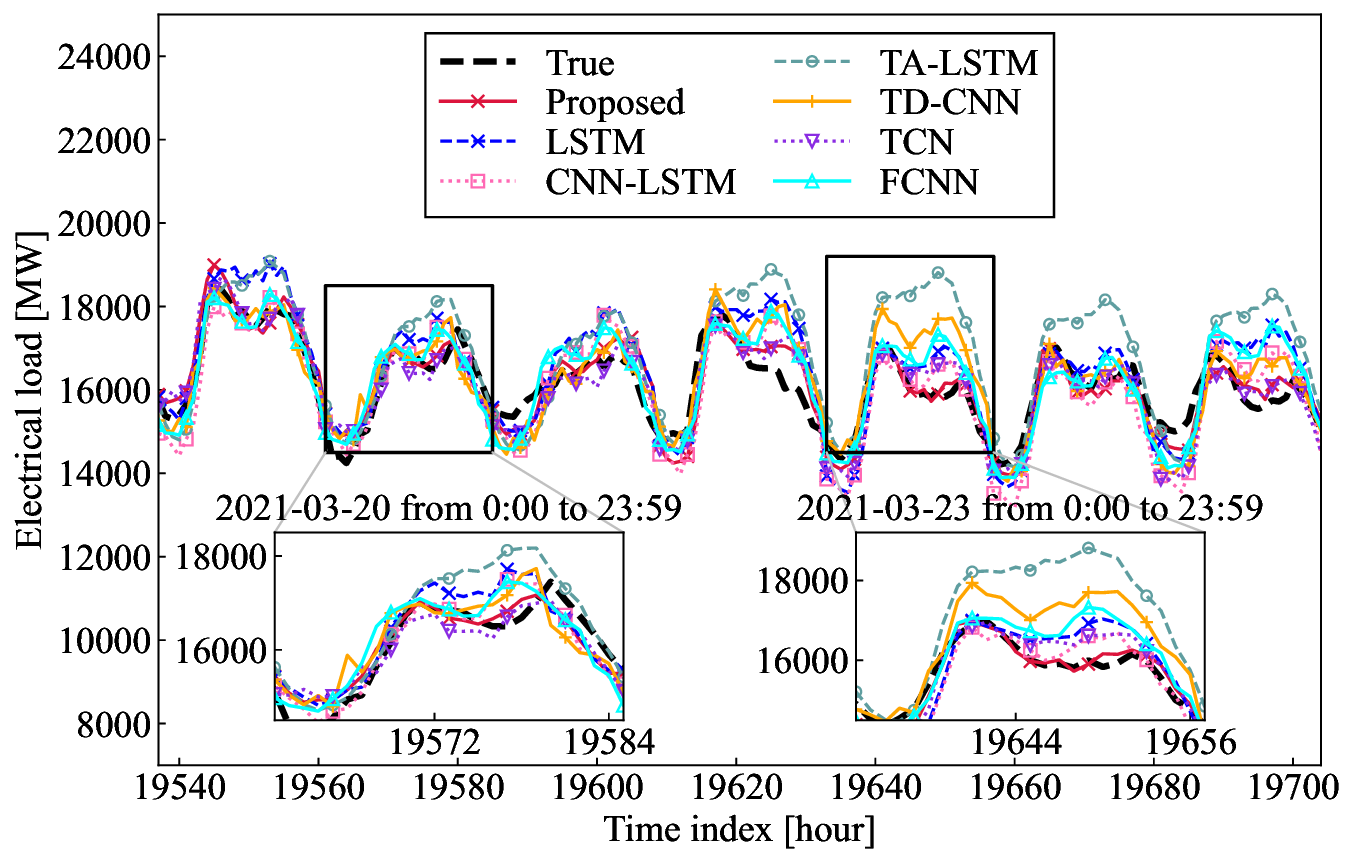}
  \caption{Day-ahead electrical load forecasting results of the proposed model and benchmarks from 2021-03-19 to 2021-03-25 in Norway.}
  \label{benchmark comparison norway}
\end{figure}

The comparison results of the proposed model and its six benchmarks are provided in Fig. \ref{benchmark comparison norway} and TABLE \ref{commonly used deep learning}. It is especially worth noting that the proposed model outperforms the best-performing benchmark method, where the MAPE indicator can be reduced by $18.19\%$ in Belgium, $18.67\%$ in Denmark, and $5.90\%$ in Norway. 

\begin{center}
	\begin{table}[!htb]
		\renewcommand\arraystretch{0.9}
		\caption{Day-ahead forecasting results in 2021 of different methods}
		\begin{center}
			\setlength{\tabcolsep}{1.8mm}{
				\begin{tabular}{llccc}
					\toprule
					\textbf{Dataset} &\textbf{Method} & \textbf{{MAE} [MW]}& \textbf{{MAPE} [\%]}& \textbf{{RMSE} [MW]}\\
					\midrule
					\multirow {7}*{Belgium}&{Proposed}& \textbf{255.718} & \textbf{2.640} & \textbf{352.044}\\
					&LSTM& 360.479 & 3.769 & 463.618\\
					&CNN-LSTM& 349.505 & 3.615 & 446.084\\
                    &TA-LSTM& 360.097 & 3.746 & 476.818\\
					&TD-CNN& 311.769 & 3.227 & 413.892\\
                    &TCN& 306.765 & 3.190 & 414.268 \\
                    &FCNN& 441.043 & 4.536 & 571.367\\
					\midrule
                    \multirow {7}*{Denmark}&{Proposed}& \textbf{42.550} & \textbf{2.692} & \textbf{59.910} \\
					&LSTM& 63.413 & 3.998 & 91.448 \\
                    &CNN-LSTM& 63.571 & 4.113 & 90.675 \\
                    &TA-LSTM& 72.056 & 4.580 & 99.747 \\
                    &TD-CNN& 58.228 & 3.629 & 80.387 \\
                    &TCN& 58.806 & 3.710 & 80.948 \\
                    &FCNN& 53.250 & 3.310 & 74.271 \\
                    \midrule
                    \multirow {7}*{Norway}&{Proposed}& \textbf{339.112} & \textbf{2.152} & \textbf{457.098} \\
					&LSTM& 538.706 & 3.449 & 708.664 \\
                    &CNN-LSTM& 603.822 & 3.667 & 815.634 \\
                    &TA-LSTM& 707.005 & 4.327 & 959.898 \\
					&TD-CNN& 531.522 & 3.310 & 700.900 \\
                    &TCN& 362.117 & 2.287 & 478.201 \\
					&FCNN& 479.018 & 2.976 & 634.240 \\
                    \bottomrule
			\end{tabular}}
		\end{center}
		\label{commonly used deep learning}
	\end{table}
\end{center}

\subsection{Comparison with Models Using Different Online Correction Strategies}
To show the effectiveness of the online model correction strategy developed based on output layer parameter fine-tuning, the comparison methods with different correction strategies are provided here:
\begin{enumerate}
\item Model 4: The model without online correction.
\item Model 5: The model with all parameters online retrained using the latest data.
\end{enumerate}
 The online correction frequency of these two methods is also set as once a week. It should be pointed out that retraining all model parameters results in a worse forecasting result than only fine-tuning the output layer parameters, which shows the importance of maintaining anti-disturbance acquired in the offline training phase. More statistical details are shown in TABLE \ref{comparison online correction}, which illustrates the proposed online correction strategy can reduce the MAPE by at least $8.71\%$, $7.30\%$, and $11.77\%$ in Belgium, Denmark, and Norway, respectively, compared to the other two strategies.

\begin{center}
	\begin{table}[!htbp]
		\renewcommand\arraystretch{0.9}
		\caption{Day-ahead forecasting results in 2021 of models with different online correction strategies}
		\begin{center}
			\setlength{\tabcolsep}{2mm}{
				\begin{tabular}{llccc}
					\toprule
					\textbf{Dataset} &\textbf{Method} & \textbf{{MAE} [MW]}& \textbf{{MAPE} [\%]}& \textbf{{RMSE} [MW]}\\
					\midrule
					\multirow {3}*{Belgium}&{Proposed}& \textbf{255.718} & \textbf{2.640} & \textbf{352.044}\\
					&Model 4& 304.968 & 3.130 & 394.753\\
					&Model 5& 278.759 & 2.892 & 382.719\\
					\midrule
                    \multirow {3}*{Denmark}&{Proposed}& \textbf{42.550} & \textbf{2.692} & \textbf{59.910} \\
					&Model 4& 45.944 & 2.904 & 64.112  \\
                    &Model 5& 49.658 & 3.143 & 68.640 \\
                    \midrule
                    \multirow {3}*{Norway}&{Proposed}& \textbf{339.112} & \textbf{2.152} & \textbf{457.098} \\
					&Model 4& 380.771 & 2.439 & 497.795 \\
                    &Model 5& 423.952 & 2.702 & 568.344 \\
                    \bottomrule
			\end{tabular}}
		\end{center}
		\label{comparison online correction}
	\end{table}
\end{center}

\section{CONCLUSION}
In this study, an advanced day-ahead electrical load forecasting method is proposed. Firstly, an intuitive and comprehensive feature selection strategy is employed to extract four types of features based on the characteristics of the electrical load time series.
Then, a hybrid LSTM-based electrical load forecasting model is designed, which is composed of an LSTM neural network block to handle the temporal feature of the historical time series and a multi-layer FCNN block to model three types of non-temporal features. Moreover, the model can remain calibrated to fit the latest electrical load distribution while retraining the anti-disturbance capability by the gradient-regularization-based offline training and online correction based on output layer parameter fine-tuning. Extensive experimental results demonstrate the superior performance of the proposed day-ahead electrical load forecasting method relative to all the considered state-of-the-art benchmarks. 

\bibliography{IEEEexample}

\begin{thebibliography}{10}
\providecommand{\url}[1]{#1}
\csname url@samestyle\endcsname
\providecommand{\newblock}{\relax}
\providecommand{\bibinfo}[2]{#2}
\providecommand{\BIBentrySTDinterwordspacing}{\spaceskip=0pt\relax}
\providecommand{\BIBentryALTinterwordstretchfactor}{4}
\providecommand{\BIBentryALTinterwordspacing}{\spaceskip=\fontdimen2\font plus
\BIBentryALTinterwordstretchfactor\fontdimen3\font minus
  \fontdimen4\font\relax}
\providecommand{\BIBforeignlanguage}[2]{{%
\expandafter\ifx\csname l@#1\endcsname\relax
\typeout{** WARNING: IEEEtran.bst: No hyphenation pattern has been}%
\typeout{** loaded for the language `#1'. Using the pattern for}%
\typeout{** the default language instead.}%
\else
\language=\csname l@#1\endcsname
\fi
#2}}
\providecommand{\BIBdecl}{\relax}
\BIBdecl

\bibitem{hobbs1999analysis}
B.~F. Hobbs, S.~Jitprapaikulsarn, S.~Konda, V.~Chankong, K.~A. Loparo, and
  D.~J. Maratukulam, ``Analysis of the value for unit commitment of improved
  load forecasts,'' \emph{IEEE Trans. Power Syst.}, vol.~14, no.~4, pp.
  1342--1348, 1999.

\bibitem{KUSTER2017257}
C.~Kuster, Y.~Rezgui, and M.~Mourshed, ``Electrical load forecasting models: A
  critical systematic review,'' \emph{Sust. Cities Soc.}, vol.~35, pp.
  257--270, 2017.

\bibitem{amjady2001short}
N.~Amjady, ``Short-term hourly load forecasting using time-series modeling with
  peak load estimation capability,'' \emph{IEEE Trans. Power Syst.}, vol.~16,
  no.~3, pp. 498--505, 2001.

\bibitem{7137662}
B.~Liu, J.~Nowotarski, T.~Hong, and R.~Weron, ``Probabilistic load forecasting
  via quantile regression averaging on sister forecasts,'' \emph{IEEE Trans.
  Smart Grid}, vol.~8, no.~2, pp. 730--737, 2017.

\bibitem{wang2020short}
Y.~Wang, J.~Chen, X.~Chen, X.~Zeng, Y.~Kong, S.~Sun, Y.~Guo, and Y.~Liu,
  ``Short-term load forecasting for industrial customers based on
  {TCN}-light{GBM},'' \emph{IEEE Trans. Power Syst.}, vol.~36, no.~3, pp.
  1984--1997, 2020.

\bibitem{en11010213}
P.-H. Kuo and C.-J. Huang, ``A high precision artificial neural networks model
  for short-term energy load forecasting,'' \emph{Energies}, vol.~11, no.~1,
  2018.

\bibitem{7748604}
H.~Jiang, Y.~Zhang, E.~Muljadi, J.~J. Zhang, and D.~W. Gao, ``A short-term and
  high-resolution distribution system load forecasting approach using support
  vector regression with hybrid parameters optimization,'' \emph{IEEE Trans.
  Smart Grid}, vol.~9, no.~4, pp. 3341--3350, 2018.

\bibitem{982201}
T.~Senjyu, H.~Takara, K.~Uezato, and T.~Funabashi, ``One-hour-ahead load
  forecasting using neural network,'' \emph{IEEE Trans. Power Syst.}, vol.~17,
  no.~1, pp. 113--118, 2002.

\bibitem{JURADO2015276}
S.~Jurado, Àngela Nebot, F.~Mugica, and N.~Avellana, ``Hybrid methodologies
  for electricity load forecasting: Entropy-based feature selection with
  machine learning and soft computing techniques,'' \emph{Energy}, vol.~86, pp.
  276--291, 2015.

\bibitem{kong2017short}
W.~Kong, Z.~Y. Dong, Y.~Jia, D.~J. Hill, Y.~Xu, and Y.~Zhang, ``Short-term
  residential load forecasting based on {LSTM} recurrent neural network,''
  \emph{IEEE Trans. Smart Grid}, vol.~10, no.~1, pp. 841--851, 2019.

\bibitem{BASHIR20221678}
T.~Bashir, C.~Haoyong, M.~F. Tahir, and Z.~Liqiang, ``Short term electricity
  load forecasting using hybrid prophet-{LSTM} model optimized by {BPNN},''
  \emph{Energy Rep.}, vol.~8, pp. 1678--1686, 2022.

\bibitem{elia}
\BIBentryALTinterwordspacing
Elia. Total load on the belgian grid. [Online]. Available:
  \url{https://opendata.elia.be/explore/dataset/ods001/information}
\BIBentrySTDinterwordspacing

\bibitem{energinet}
\BIBentryALTinterwordspacing
Energinet. Production and consumption - settlement. [Online]. Available:
  \url{https://www.energidataservice.dk/tso-electricity/ProductionConsumptionSettlement}
\BIBentrySTDinterwordspacing

\bibitem{statnett}
\BIBentryALTinterwordspacing
Statnett. Data from the power system. [Online]. Available:
  \url{https://www.statnett.no/en/for-stakeholders-in-the-power-industry/data-from-the-power-system}
\BIBentrySTDinterwordspacing

\bibitem{zhi2021con}
Z.~Lv, H.~Ding, L.~Wang, and Q.~Zou, ``A convolutional neural network using
  dinucleotide one-hot encoder for identifying dna n6-methyladenine sites in
  the rice genome,'' \emph{Neurocomputing}, vol. 422, pp. 214--221, 2021.

\bibitem{DUDEK2015277}
G.~Dudek, ``Pattern similarity-based methods for short-term load forecasting
  – part 1: Principles,'' \emph{Appl. Soft. Comput.}, vol.~37, pp. 277--287,
  2015.

\bibitem{HUANG20211041}
X.~Huang, Q.~Li, Y.~Tai, Z.~Chen, J.~Zhang, J.~Shi, B.~Gao, and W.~Liu,
  ``Hybrid deep neural model for hourly solar irradiance forecasting,''
  \emph{Renew. Energy}, vol. 171, pp. 1041--1060, 2021.

\bibitem{MA2022123217}
T.-Y. Ma and S.~Faye, ``Multistep electric vehicle charging station occupancy
  prediction using hybrid {LSTM} neural networks,'' \emph{Energy}, vol. 244, p.
  123217, 2022.

\bibitem{Neco1997}
S.~Hochreiter and J.~Schmidhuber, ``{Long Short-Term Memory},'' \emph{Neural
  Comput.}, vol.~9, no.~8, pp. 1735--1780, 11 1997.

\bibitem{PHAM20181}
D.-H. Pham and A.-C. Le, ``Exploiting multiple word embeddings and one-hot
  character vectors for aspect-based sentiment analysis,'' \emph{Int. J.
  Approx. Reasoning}, vol. 103, pp. 1--10, 2018.

\bibitem{schwing2015fully}
A.~G. Schwing and R.~Urtasun, ``Fully connected deep structured networks,''
  \emph{arXiv preprint arXiv:1503.02351}, 2015.

\bibitem{ross2018improving}
A.~Ross and F.~Doshi-Velez, ``Improving the adversarial robustness and
  interpretability of deep neural networks by regularizing their input
  gradients,'' in \emph{Proceedings of the AAAI Conference on Artificial
  Intelligence}, vol.~32, no.~1, 2018.

\bibitem{miyato2016adversarial}
T.~Miyato, A.~M. Dai, and I.~Goodfellow, ``Adversarial training methods for
  semi-supervised text classification,'' \emph{arXiv preprint
  arXiv:1605.07725}, 2016.

\bibitem{kingma2014adam}
D.~P. Kingma and J.~Ba, ``Adam: A method for stochastic optimization,''
  \emph{arXiv preprint arXiv:1412.6980}, 2014.

\bibitem{li2018learning}
Z.~Li and D.~Hoiem, ``Learning without forgetting,'' \emph{IEEE Trans. Pattern
  Anal. Mach. Intell.}, vol.~40, no.~12, pp. 2935--2947, 2018.

\bibitem{kim2019predicting}
T.-Y. Kim and S.-B. Cho, ``Predicting residential energy consumption using
  {CNN-LSTM} neural networks,'' \emph{Energy}, vol. 182, pp. 72--81, 2019.

\bibitem{zang2021residential}
H.~Zang, R.~Xu, L.~Cheng, T.~Ding, L.~Liu, Z.~Wei, and G.~Sun, ``Residential
  load forecasting based on {LSTM} fusing self-attention mechanism with
  pooling,'' \emph{Energy}, vol. 229, p. 120682, 2021.

\bibitem{han2018enhanced}
L.~Han, Y.~Peng, Y.~Li, B.~Yong, Q.~Zhou, and L.~Shu, ``Enhanced deep networks
  for short-term and medium-term load forecasting,'' \emph{IEEE Access},
  vol.~7, pp. 4045--4055, 2018.

\bibitem{zhu2020short}
R.~Zhu, W.~Liao, and Y.~Wang, ``Short-term prediction for wind power based on
  temporal convolutional network,'' \emph{Energy Rep.}, vol.~6, pp. 424--429,
  2020.

\bibitem{ryu2016deep}
A.~Ahmad, N.~Javaid, M.~Guizani, N.~Alrajeh, and Z.~A. Khan, ``An accurate and
  fast converging short-term load forecasting model for industrial applications
  in a smart grid,'' \emph{IEEE Trans. Ind. Informat.}, vol.~13, no.~5, pp.
  2587--2596, 2017.

\end{thebibliography}
\bibliographystyle{IEEEtran}


\end{document}